# Ultrafast Electron Microscopy:

# A Quantitative Platform for Nonequilibrium Materials Research

David J. Flannigan* and Swarit Ahmed Shadman

*Department of Chemical Engineering and Materials Science, University of Minnesota, 421 Washington Avenue SE, Minneapolis, MN 55455, USA*

**Abstract:** Macroscopic materials function is determined not merely by equilibrium structure, but by how carriers, phonons, fields, defects, interfaces, and collective order dynamically evolve after perturbation. Ultrafast electron microscopy (UEM) uniquely bridges this gap, coupling femtosecond-to-nanosecond timing with real-space, reciprocal-space, and energy-resolved contrast. Here, we review how these integrated capabilities now quantitatively map energy flow and conversion in electronic materials, decode electronic-structural coupling in quantum and correlated systems, and isolate the localized propagation of carriers, polaritons, strain, and phonons across optoelectronic and nanomechanical architectures. We emphasize the predictive design rules now emerging from direct metrology of carrier-lattice coupling, momentum-resolved phonon thermalization, defect-controlled phase transformations, and authentic *operando* switching. We critically assess UEM's ongoing maturation from a qualitative, proof-of-concept experiment into a rigorous, quantitative materials platform. This paradigm shift is actively driven by integrated multimodal detection, advanced energy-resolved spectroscopy, high-repetition-rate sources, multidimensional (4D and 5D) acquisition, and physically constrained forward modeling. Ultimately, the central opportunity lies in establishing a complete, causal link from initial excitation to localized energy flow, transient structure, and macroscopic device function. By prioritizing quantitative reproducibility and robust mechanistic interpretation, UEM is positioned to engineer materials in functional states that equilibrium thermodynamics simply cannot describe.

*Corresponding author
Email: flan0076@umn.edu
Office: +1 612-625-3867

## 1. Introduction

Functional materials driven by light, electrical bias, or other stimuli often evolve through nonequilibrium, sometimes spatially heterogeneous states inaccessible to single-equilibrium structural descriptions (1–4). Consequently, functional responses depend on how charge carriers, phonons, defects, interfaces, and collective order parameters exchange energy after excitation. For example, in photovoltaic absorbers, key questions involve local polaronic distortions, excited phonon modes, ion migration, and interfacial extraction-recombination kinetics (5–9). In correlated materials, lattice responses are typically material-, fluence-, strain-, and geometry-dependent rather than following a universal sequence of events (4, 10–14). Similarly, the performance of nanomechanical and optoelectronic devices can depend on propagating strain waves, free carriers, and polariton wave packets rather than spatially-averaged ensemble behaviors (15–18).

Resolving these nonequilibrium dynamics requires measurement capabilities that are matched across spatial, energy, momentum, and temporal scales (1, 19). Far-field optical pump-probe spectroscopies provide high temporal resolution and sensitivity to carrier populations and collective excitations, but they inherently spatially average over domains, grains, interfaces, and competing pathways – a limitation only partially mitigated by ultrafast optical nanoimaging (20). Conventional transmission electron microscopy (TEM) delivers atomic-to-nanoscale imaging, diffraction, and spectroscopy (21), yet detectors under continuous-beam operation cannot resolve femtosecond-to-picosecond bond rearrangements or phonon propagation without temporally structured electron pulses synchronized to a clocking excitation (19, 22, 23). While ultrafast X-ray scattering provides exceptional structural sensitivity, deep penetration, and minimal multiple scattering, the larger elastic scattering cross section of energetic electrons provides a distinct advantage for probing thin, low-dimensional, and small-volume specimens, provided the sample can be made sufficiently thin and multiple/dynamical scattering is modeled (24–27).

By spanning imaging, diffraction, near-field, and energy-loss modalities, ultrafast electron microscopy (UEM) accesses these spaces in complementary, differently-optimized modes (Figure 1) (19, 28). However, these modalities impose fundamental trade-offs among spatiotemporal resolution, energy spread, beam current, dose, and signal-to-noise ratio (22, 23, 28–30). Importantly, a time-resolved intensity change rarely provides a unique measure of the underlying material state; diffraction and bright-field intensities are reshaped by dynamical scattering and excitation error, while sub-degree tilts or nanometer translations can mimic Debye-Waller attenuation or structural motion (31–33). Further, low-current pulsed beams produce weak counting statistics susceptible to drift, contamination, acquisition-order artifacts, and cumulative heating if incomplete specimen recovery occurs between pump cycles (22, 34–37). Therefore, rigorous assignment requires triangulating multiple constraints – such as combining time-resolved diffraction with electron energy-loss spectroscopy (EELS), energy-filtered imaging with core-loss spectroscopy, or diffuse scattering with lattice-dynamical calculations (29, 31, 38, 39). Establishing this across all UEM measurements is attractive, as distinguishing true dynamics from artifacts or specimen-specific properties (*e.g.*, geometry and thickness) remains an interpretive challenge.

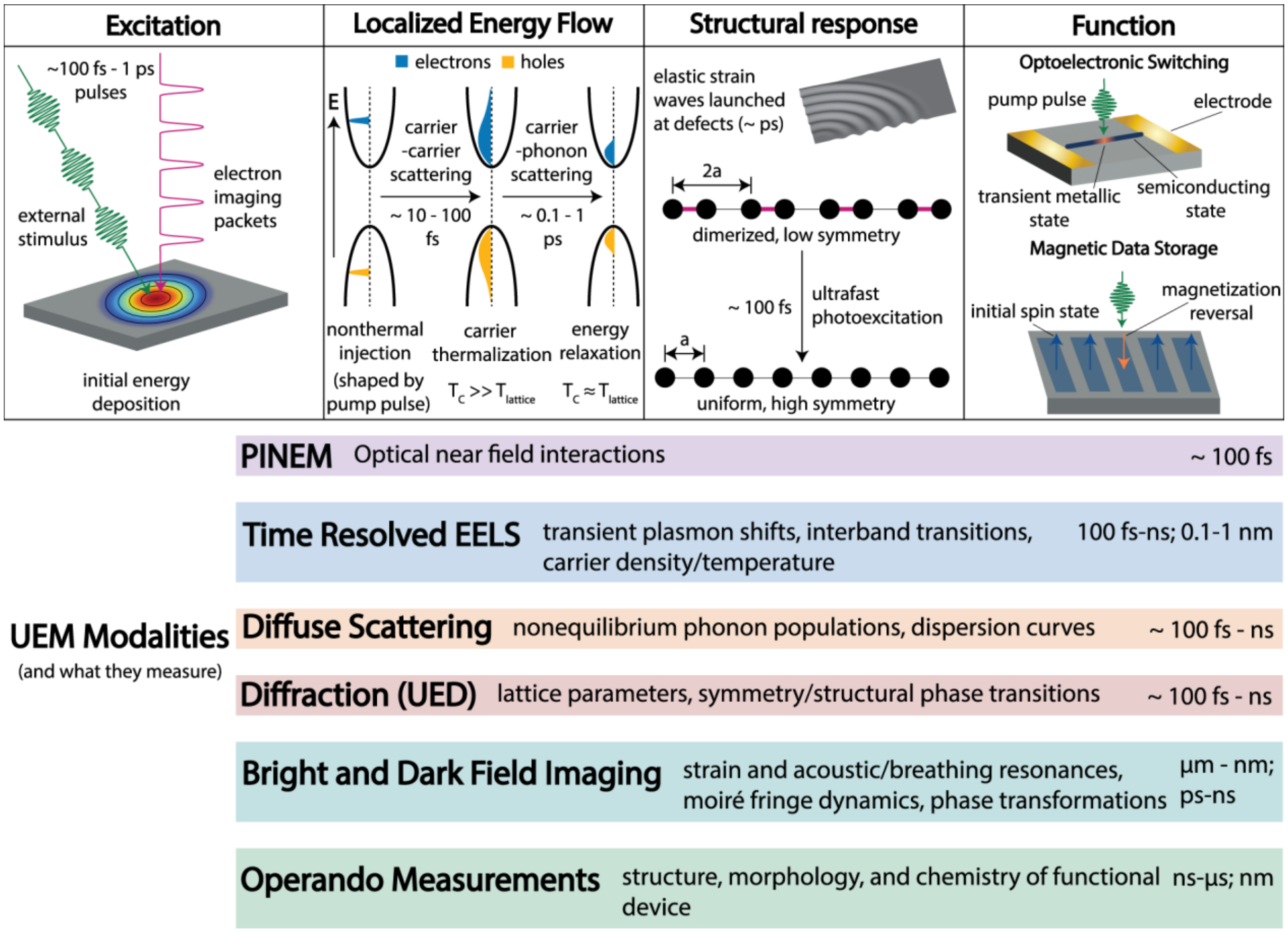

**Figure 1. UEM as an energy-to-structure-to-function probe.** The four upper panels – Excitation, Localized Energy Flow, Structural Response, and Function – illustrate the general nonequilibrium materials heirarchical energy-flow paradigm targeted by UEM interrogation. Optical (typically) excitation combined with electron-packet probing excites carriers locally, with scattering, thermalization, and relaxation evolving over space and time. Nonequilibrium carrier populations and density gradients produce lattice responses such as transient strain, coherent phonons, and changes in symmetry. Combined with *in situ* and *operando* methods, UEM is increasingly being used to probe real-time device function. We argue that one of the main benefits of UEM is the myriad (differently-optimized) modalities combined within one properly-equipped

instrument [*e.g.*, photon-induced near-field electron microscopy (PINEM), ultrafast EELS, imaging, *etc.*]. Figure layout and concepts informed by (1, 2, 19, 29, 40).

Selected *operando* and correlative studies are moving UEM beyond nominal femtosecond contrast toward calibrated and statistically-auditable measurements (34, 41, 42). Further, ultrafast imaging of transient morphologies has been augmented by laser-driven cold field-emission sources that deliver high coherence, sub-eV energy spreads, and angstrom-scale focusability (23, 30, 34, 43–45). Concurrent instrumental advances – including lossless monochromation for low-loss phonon and exciton spectroscopy, megahertz (MHz) cycling on thermally engineered supports, four-dimensional precession electron diffraction (4D-PED), and electrically pumped *operando* switching – are now bridging the gap to potentially connect localized structural and electronic responses directly to macroscale device functionality in selected systems (29, 31, 35, 38, 40–42, 46–49).

Here, we examine how UEM can be utilized to trace the redistribution of deposited energy among electronic, structural, and collective degrees of freedom to link these pathways to measurable materials outcomes. Following foundational overviews of the basic instrumentation (Section 2) and uniquely-resolved UEM capabilities and physical constraints (Section 3), we evaluate critical nonequilibrium pathways across distinct functional domains: carrier-lattice thermalization in energy materials (Section 4), hidden order and competing degrees of freedom in quantum and correlated materials (Section 5), localized carrier and polariton transport in optoelectronics, and defect-launched coherent elastic waves in nanomechanical materials (Section 6). Finally, we assess UEM's maturation from a high-speed imaging tool into a quantitative, multimodal materials platform (Section 7) and outline the paradigm shift this enables for nonequilibrium materials science and engineering (Section 8).

## 2. Fundamentals of UEM Instrumentation

To properly contextualize UEM applications, it is important to first briefly define the fundamental architecture and inherent limitations of the instruments. The basic layout of a laser-driven stroboscopic UEM (the type of instrument mainly focused on here) relies on a synchronized optical-pump/electron-probe configuration (Figure 2) (23, 40, 50–53). A femtosecond laser pulse is directed onto the specimen to initiate a nonequilibrium state, while a delayed, ultrashort electron pulse is generated and accelerated down the column to probe the transient response. Operating in this stroboscopic transmission mode imposes strict sample requirements: specimens must be electron-transparent while possessing the thermal resilience and structural stability to produce a reprobucible, response at the chosen repetition rate (22, 34, 35, 54). Furthermore, the technique faces inherent resolution limits driven by Coulombic repulsion (space-charge effects) within the discrete electron packets, which degrades spatial coherence, energy spread, and temporal resolution at high electron densities (55–58).

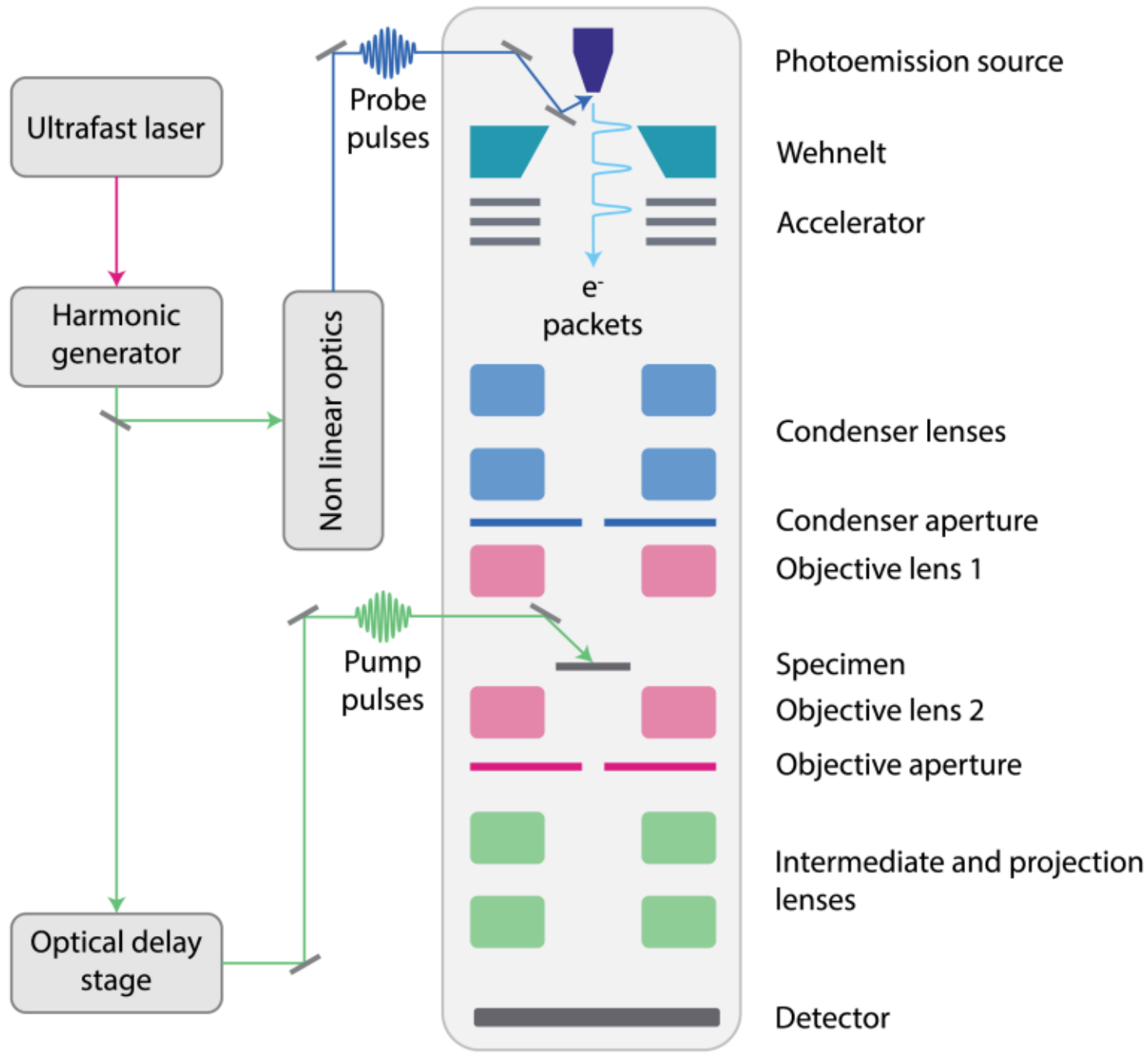

**Figure 2. Basic layout of a stroboscopic UEM instrument.** Shown are the essential components configured for ultrafast transmission electron microscopy operation. Pulses from an ultrafast laser source undergo harmonic conversion before being split into two separate beamlines. One, in this case the pump-pulse line, enters an optical delay stage used to vary the relative arrival times of photoelectrons and excitation photons at the specimen. The second typically undergoes a second harmonic conversion to UV pulses and is used to generate discrete photoelectron packets at the TEM photoemission source. The base TEM configuration for UEMs constructed from commercial instruments can either be largely unchanged or can be modified with an additional condenser lens (not shown) to gain direct control over the intial crossover behavior in the electron gun region.

The architecture of the pulsed electron gun fundamentally dictates the spatial coherence and energy resolution achievable across various TEM modes. Many of the foundational stroboscopic UEM studies relied on flat, truncated photocathodes (51, 54, 59–62). While these designs reliably generate high-density electron packets, their large emission areas typically result in lower beam brightness and broader energy spreads (63). Advancements have been made through the integration of sharp field emitters, including Schottky and cold-field-emission tips, into UEM platforms . These sharp emitters provide higher beam brightness, improved spatial coherence, and also lower energy spreads, depending upon the specific emission regime and operating conditions (*e.g.*, emitter temperature, photoemission conditions, extractor fields, laser parameters, and bunch charge) (23, 30, 44, 45, 64–66). Indeed, specific platforms have demonstrated sub-nanometer or even Å-scale probes, but these benchmarks should not be generalized to all UEM platforms. This evolution in gun design is an important factor in pushing the high-resolution capabilities of advanced modalities like 4D-STEM and ultrafast EELS.

In addition to photocathode-driven pulsed electron sources, ultrafast beam blankers provide an effective mechanism for achieving required temporal resolutions (67–70). By rapidly deflecting a continuous electron beam across a chopping aperture using radio-frequency (RF) or electrostatic fields, modern blankers can carve out picosecond electron pulses without requiring a pulsed drive laser at the gun. Integrating ultrafast beam blankers into the UEM column opens new possibilities for stroboscopic imaging, particularly by facilitating the coupling of non-laser-induced excitations with the electron pulses. This capability is important for potentially advancing authentic *operando* architectures (*i.e.*, measurements performed under realistic device operating conditions), as it allows researchers to synchronize the electron probe with rapid electrical biasing, cyclic mechanical stress, or direct thermal transients, thereby expanding the technique beyond traditional optical-pump paradigms (41, 67, 71).

Because femtosecond-picosecond UEM relies on stroboscopic acquisition – building sufficient signal-to-noise ratios across thousands or millions of identical pump-probe cycles – its application is fundamentally limited to repeatable or stationary periodic phenomena. The material must return to its initial state, which need not necessarily be its ground state, within the period between sequential electron pulses so that a single unique state is probed at each delay time. This stroboscopic requirement precludes the direct observation of irreversible processes, such as permanent phase transformations, material fracture, or cumulative structural degradation. Capturing such single-event, irreversible phenomena requires the high-current, single-shot capabilities of dynamic or movie-mode TEM (DTEM) or other related non-stroboscopic pulsed TEM methods (53, 72–79). While we focus here on stroboscopic, laser-driven stroboscopic UEM owing to space constraints (with additional mention of scanning UEM when discussing carrier dynamics), we emphasize that complementary modalities – dynamic and movie-mode TEM and chopped-beam UEM – are rich in materials-research applications and device physics and are thus, in our view, well-deserving of standalone accounts.

## 3. What UEM Resolves

The distinctive value of UEM lies not solely in an absolute temporal or spatial resolution advantage, but in its capacity to correlate real-space imaging, reciprocal-space diffraction, energy analysis, and field-sensitive contrast within a presurveyed nanoscale region (40). By interrogating this region following a synchronized excitation, structural, electronic, and electromagnetic responses can be evaluated within a common spatial frame (28, 52, 80). This triangulation of modalities constrains nonequilibrium pathways by identifying response ordering, localized nucleation sites, and transient-state correlations. However, establishing definitive causal and functional assignments invariably requires complementary measurements and physical modeling to navigate the inherent trade-offs among spatiotemporal resolution, electron dose, energy spread, and signal-to-noise ratios (11, 27, 34, 81). The complementary constraints, physical observables, and performance trade-offs of UEM modalities are consolidated in Table 1.

Table 1. UEM Modalities, Physical Observables, and Performance Trade-offs

| **Modality, Technique, Workflow** | **Representative Demonstrations** | **Primary Physical Observable** | **Representative Material System & Insights** | **Key Trade-offs & Limitations** | **References** |
|---|---|---|---|---|---|
| Diffraction/CBED/ UKD/4D-PED | 28 nm/700 fs (CBED); sub-ps to ns | Bragg-spot positions, intensities, widths, and symmetry; diffuse scattering; local deformation tensors | $VO_2$ (photoinduced M1-M2-R intermediate state); $VTe_2$ (4D precession structural refinement); graphite & Si (nanoscale shear and acoustic waves) | Sensitive to dynamical scattering, excitation error, and specimen tilt/translation; spatially averages illuminated region | (31, 54, 82–85) |
| Bright-Field/Dark-Field/Lorentz Imaging | ~5 nm/530 fs (DF CDW); nanoscale to mesoscale | Domain conversion, acoustic/strain wavefronts, defect pinning, magnetization fields | 1*T*-$TaS_2$ (binary CDW domain growth); Ge/$WSe_2$ (step-edge phonon nucleation); $MoS_2$ (strain-wave interference); GaAs & MnNiSn (defect pinning) | BF contrast is not a direct displacement map (bend contours, orientation effects); DF intensity depends on thickness and domain fraction | (12, 16, 17, 86–90) |
| PINEM/CDEM/Time-Resolved EELS | Nanometer/~50 to 155 fs (PINEM); sub-ps (EELS/CDEM) | Quantized electron-photon exchange (evanescent fields), THz field amplitude/phase, plasmons, core-loss bonding | $VO_2$ nanowire (155 fs dielectric metallization before lattice motion); *h*-BN (polariton wave packets); InAs (carrier reconstruction); hematite ($Fe^{4+}$ oxidation) | Low-loss excitations delocalized beyond probe; weak core-loss cross sections; low brightness; time-zero broadening | (10, 18, 21, 29, 38, 46, 87, 91–95) |
| Correlative & *Operando* UEM | Nanoscale/picosecond to microsecond | Synchronized electrical/optical bias coupled to structural phase fronts and switching kinetics | Electrically driven $VO_2$ (Mott-Peierls structural switching limits); $VO_2$ nanoparticles (4D imaging + diffraction correlation) | Requires highly reproducible cycling, stable synchronization, and complete thermal recovery between excitation pulses | (11, 13, 29, 31, 38, 41, 49, 54, 85, 96) |

UEM diffraction techniques – including selected-area, convergent-beam (CBED), Kikuchi (UKD), and four-dimensional precession modes (4D-PED) – can quantify transient reciprocal-space changes. Under calibrated conditions for thickness, orientation, excitation error, and dynamical scattering, these techniques can constrain atomic displacements, isolate local deformation tensors, and extract out-of-plane expansion or shear-rotation dynamics (31, 82–85, 97). By localizing reciprocal-space probes, UEM separates localized nanoscale deformation from spatially-averaged vibrational modes (31, 98, 99). Accurate separation and quantification of these dynamics, however, depend on stable reference states, known sample geometries, and integration with finite-element or lattice-dynamical modeling (83, 84, 98, 100).

While conventional selected-area diffraction averages over an illuminated region, real-space UEM imaging establishes where responses nucleate, propagate, or become pinned (12, 16, 17, 41, 86–89, 101). Dark-field imaging, utilizing tailored satellite apertures, isolates specific structural order parameters to directly visualize nanoscale domain growth, boundary sharpening, and defect-pinned phase transitions (12). Similarly, bright-field and Lorentz imaging modes resolve the morphology-controlled propagation of acoustic wavefronts, strain-wave interference, phase fronts, and transient magnetic textures (16, 17, 86–89, 102, 103). A critical challenge in real-space imaging remains distinguishing true structural displacement from confounding contrast changes induced by bend contours, specimen thickness variations, and orientation effects (33).

In regimes where electronic and lattice responses decouple, modalities such as photon-induced near-field electron microscopy (PINEM), charge-dynamics electron microscopy (CDEM), and ultrafast EELS provide access to electron-field coupling, energy-loss, and deflection observables (10, 18, 21, 29, 38, 46, 87, 91–95, 104, 105). These techniques map quantized electron-photon exchanges, local optical near fields, THz field amplitudes, and core-loss bonding signatures. From these measurements, dielectric responses, electromagnetic fields, carrier distributions, and transient chemical states can be inferred (10, 18, 38, 46, 91, 94, 106). Such inferences are strengthened by electrodynamic modeling and transport-model inversions, which should be buttressed by detailed boundary conditions and field-to-charge calibrations (92).

Finally, UEM methodology increasingly demands the triangulation of complementary observables alongside forward modeling and external diagnostics (27, 34). Correlative imaging, diffraction, and spectroscopy help distinguish genuine structural transitions from artifacts caused by dynamical scattering, specimen tilt, and drift (27, 31, 38). Furthermore, emerging *operando* UEM platforms integrate synchronized optical or electrical biasing to track nanosecond-to-microsecond phase nucleation and structural switching (11, 13, 41, 49). Importantly, structural phase-front motion and electrical switching should not be treated as identical observables; their timing can diverge due to thermal dissipation, contact geometry, filament formation, and metastable phase coexistence (11, 41, 107). Indeed, multimodal agreement alone is insufficient to establish structure-function relationships; electron contrast must be calibrated against geometry, dynamical scattering, pulse timing, dose, and thermal recovery, ideally with interleaved non-driven controls (27, 34, 41). Figure 3 roughly maps the contrast-modality space, with approximate timescales and materials properties noted for each modality.

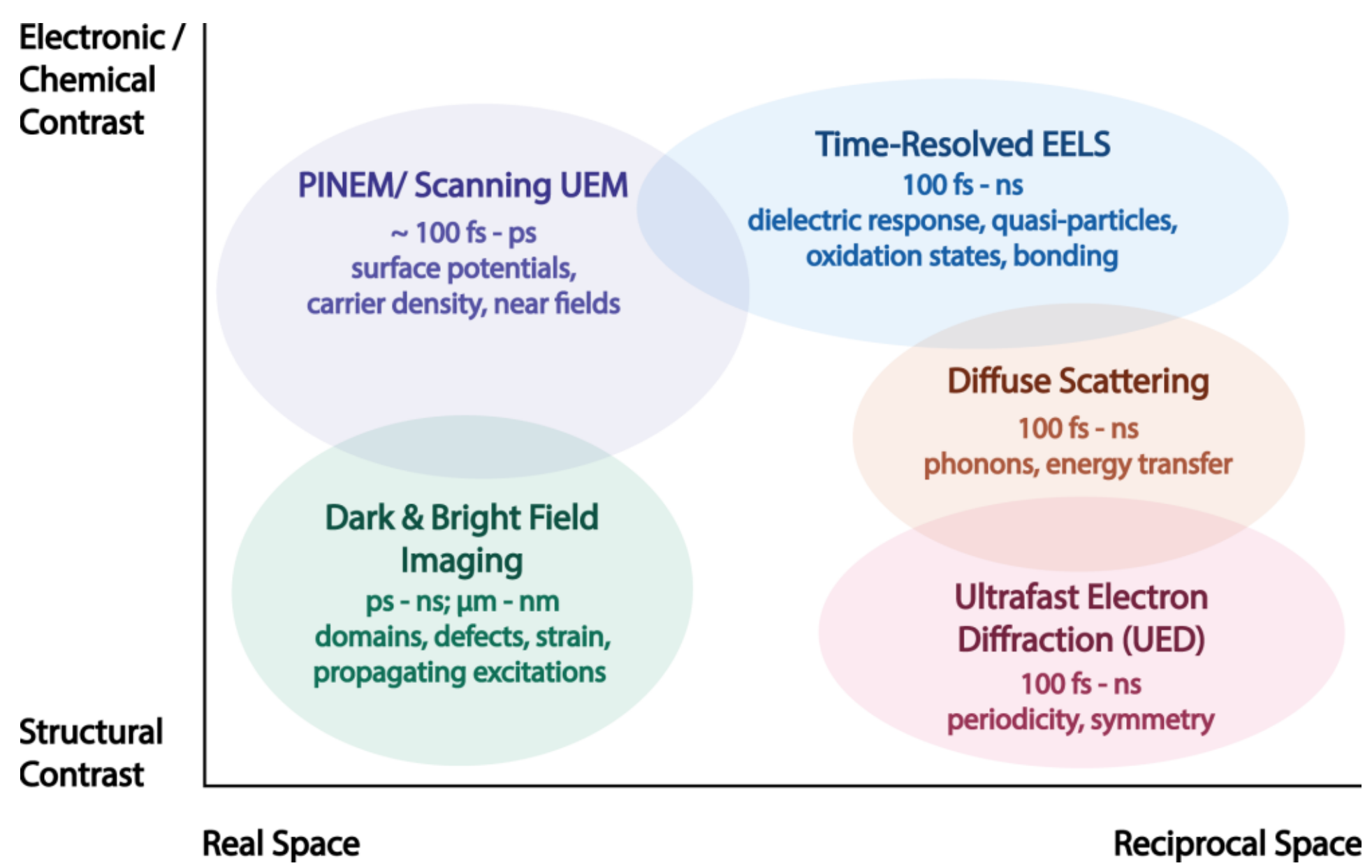

**Figure 3. Complementary information carried by ultrafast electron modalities comprising UEM.** The horizontal axis represents real-space to reciprocal-space modalities, and the vertical axis represents structural to electronic/chemical contrast. Approximate timescales and materials properties accessible to the associated modality are shown. Some modality regions overlap because some accessible information is common to each; boundaries mark shared territory rather than strict instrumental limits and should not be read as thresholds. Because no single region spans both axes, a complete materials interpretation often requires two or more observables acquired within a common frame, which is the argument developed here. Figure layout and content informed specifically by (1, 19, 21, 29).

4. Energy Materials: From Carrier Excitation to Device Function

At the device level, the function of energy materials is governed by competition among carrier extraction, recombination, and heat dissipation. For example, in photovoltaic and photocatalytic absorbers, spatially varying cooling, trapping, recombination, and transfer across junctions can alter the fraction of photocarriers available for extraction (108, 109). In thermoelectrics, phonon dispersion, propagation, and scattering influence lattice thermal transport (110). In nanostructured semiconductors, surfaces, defects, morphology, and local structural deformation can redirect excitation toward carrier extraction, trapping, recombination, and lattice motion (16, 111–113). Spanning this diverse set of dynamics, ultrafast studies have shown that nominally similar energy materials can follow distinct early-time pathways depending on excitation conditions, morphology, carrier localization, surface states, and band offsets (6, 108, 109, 114–116). Importantly, these observables are not universally-unique energy-flow responses: scanning UEM (SUEM) secondary-electron contrast can contain contributions from hot-carrier energy distributions as well as surface photovoltage, while UEM structural contrast can depend on specimen orientation, dynamical scattering, drift, and other instrument factors (33, 34, 117). In this section, we explore UEM's current and potential future impact on understanding microscopic transport dynamics in energy materials.

### 4.1. Exciton-Polaron Formation and Carrier-Lattice Coupling

The lead-halide perovskites – a particularly timely class of materials – present an interesting, albeit challenging, potential test case for UEM. The main reason for this is that polaronic lattice distortion, hot-carrier cooling, dynamic screening, and structural disorder can generate overlapping band-edge bleaches, spectral shifts, linewidth changes, sub-gap features, and mobility transients. The interesting challenge is that these behaviors can represent distinct microscopic states not uniquely separable from optical data alone (5, 6, 118–121).

For UEM, direct structural identification of exciton-polaron formation or exciton-driven symmetry breaking has not yet been demonstrated, though dedicated diffraction and spectroscopy studies have shown success (Figure 4) (5, 6, 119). In principle, however, UEM provides a potentially enabling platform for developing measurements capable of delineating the microscopic excited states by combining high-resolution imaging with highly-localized diffraction and spectroscopy (23, 28, 52, 86). Indeed, when applied to the lead-halide perovskites (under strict dose-monitored and environmental conditions to mitigate structural degradation), these complementary modalities could provide independent constraints on local structural distortions and electronic dynamics. The reason being that, individually, the modalities have successfully mapped transient strain, deformation tensors, bonding, and oxidation-state changes in other materials, thus already providing the components needed to parse the excited states (38, 46, 82, 84, 91).

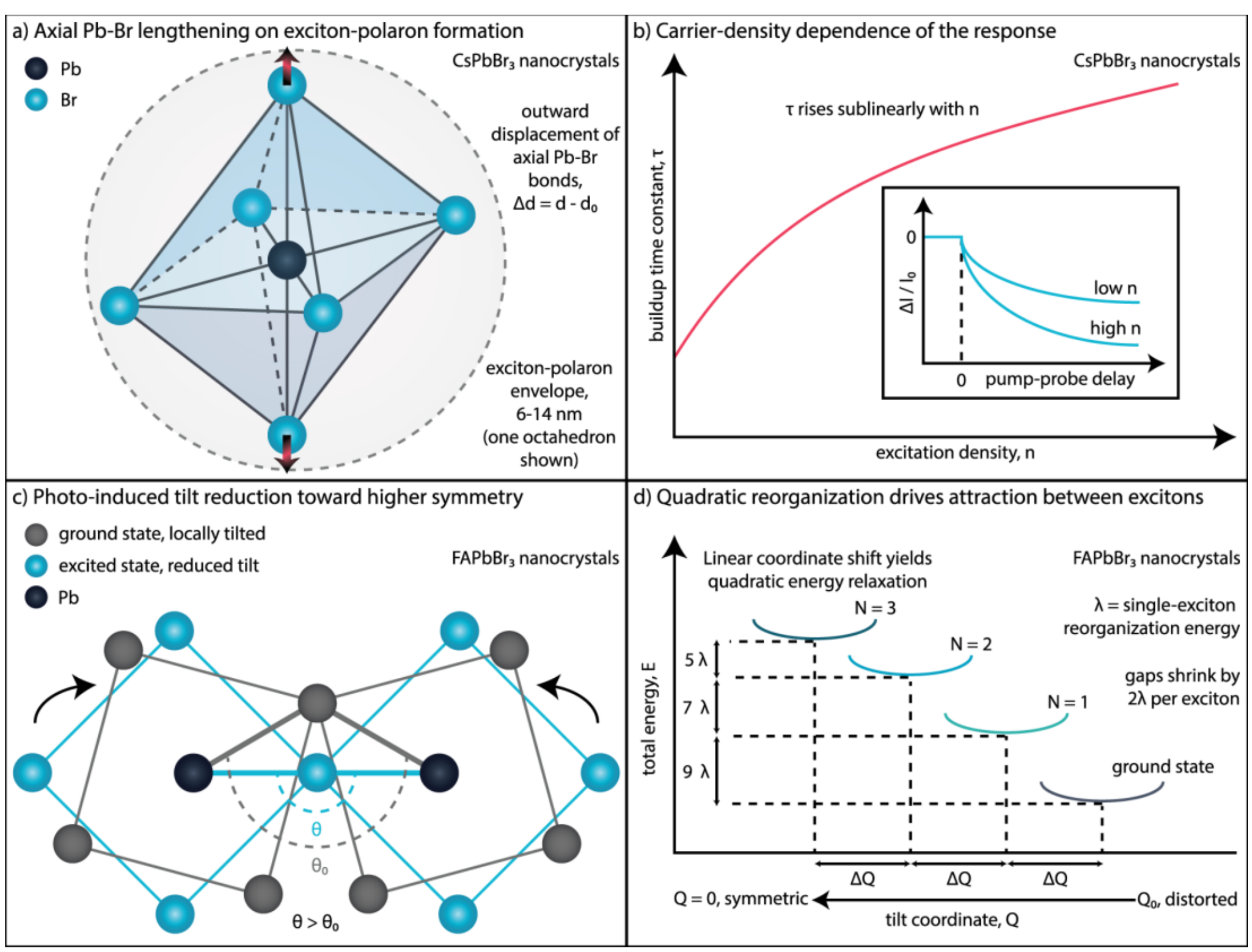

**Figure 4. The lead-halide perovskites as a state-of-the-art UEM test case.** (a) Schematic representation of photoinduced distortion of the lead bromide octahedral cage in $CsPbBr_3$, as resolved with femtosecond electron diffraction combined with structure-factor analysis and simulations. (b) Example of the information extracted from femtosecond electron diffraction for $CsPbBr_3$ dynamics: a sublinearly increasing time constant was observed with increasing photoexcitation fluence (*i.e.*, with increasing free carrier density, $n$). This was attributed to slower sub-picosecond hot-carrier cooling at higher fluence, which delays establishment of the exciton-polaron population, rather than to a hot-phonon bottleneck lasting several picoseconds. (c) Schematic representation of photoinduced lead bromide octahedral tilting resolved with in $FAPbBr_3$ nanocrystals resolved with ultrafast electron diffraction. (d) A schematic of the observed quadratic reorganization driving exciton attraction in $FAPbBr_3$. Because every exciton displaces the same tilt coordinate, individual reorganizations interfere constructively. The coupling therefore scales with the square of the exciton number. Successive multi-exciton states are thus separated by gaps that shrink by $2\lambda$ per added exciton, where $\lambda$ is the single-exciton reorganization energy. Figure content informed by (6) in panels (a,b) and by (114) in panels (c,d).

Accordingly, the closest UEM precedents are correlative real-space/diffraction measurements and local CBED strain mapping, though in non-perovskite materials (82, 84). Because independent structural studies of the hybrid perovskites have shown that static octahedral rotations vary significantly across individual nanocrystals (122, 123), future UEM experiments could combine ultrafast imaging with localized ultrafast diffraction to test whether photoinduced distortions are confined to particular defects or instead are coupled to propagating lattice modes (16, 82). Indeed, as described in Section 3, several component UEM measurements needed to study exciton-polaron formation and exciton-driven symmetry breaking in the hybrid perovskites have been established.

### 4.2. Electron-Lattice Thermalization and Nanoscale Energy Transport

Rather than providing direct measurements of thermal conductivity, UEM enables localized access to carrier-induced structural relaxation, coherent strain-wave propagation, mode conversion, and defect- or interface-dependent phonon dynamics (16, 17, 33, 82, 88, 98, 124–126). These measurements generally track structural proxies for energy flow rather than directly measuring heat flux or thermal conduction (127). For example, Figure 5 illustrates how ultrafast electron diffuse scattering in a UEM instrument was used to reveal non-thermal responses in SnSe (42). One of the few estimates of electron-lattice coupling from UEM measurements calibrated transient diffraction against static heating in nanocrystalline Au films on SiN (125). Combined with a two-temperature model, this sample-specific analysis inferred an effective coupling coefficient of $g = 7.2 \times 10^{15}\ W \cdot m^{-3} \cdot K^{-1}$. Note that differing responses in the diffraction data cautioned against treating this value as a geometry-independent material constant. Indeed, key boundary conditions – some of which are sample dependent – include film thickness, substrate type, excitation fluence, static-heating calibration, assumed two-temperature distributions, and thermal boundary conditions.

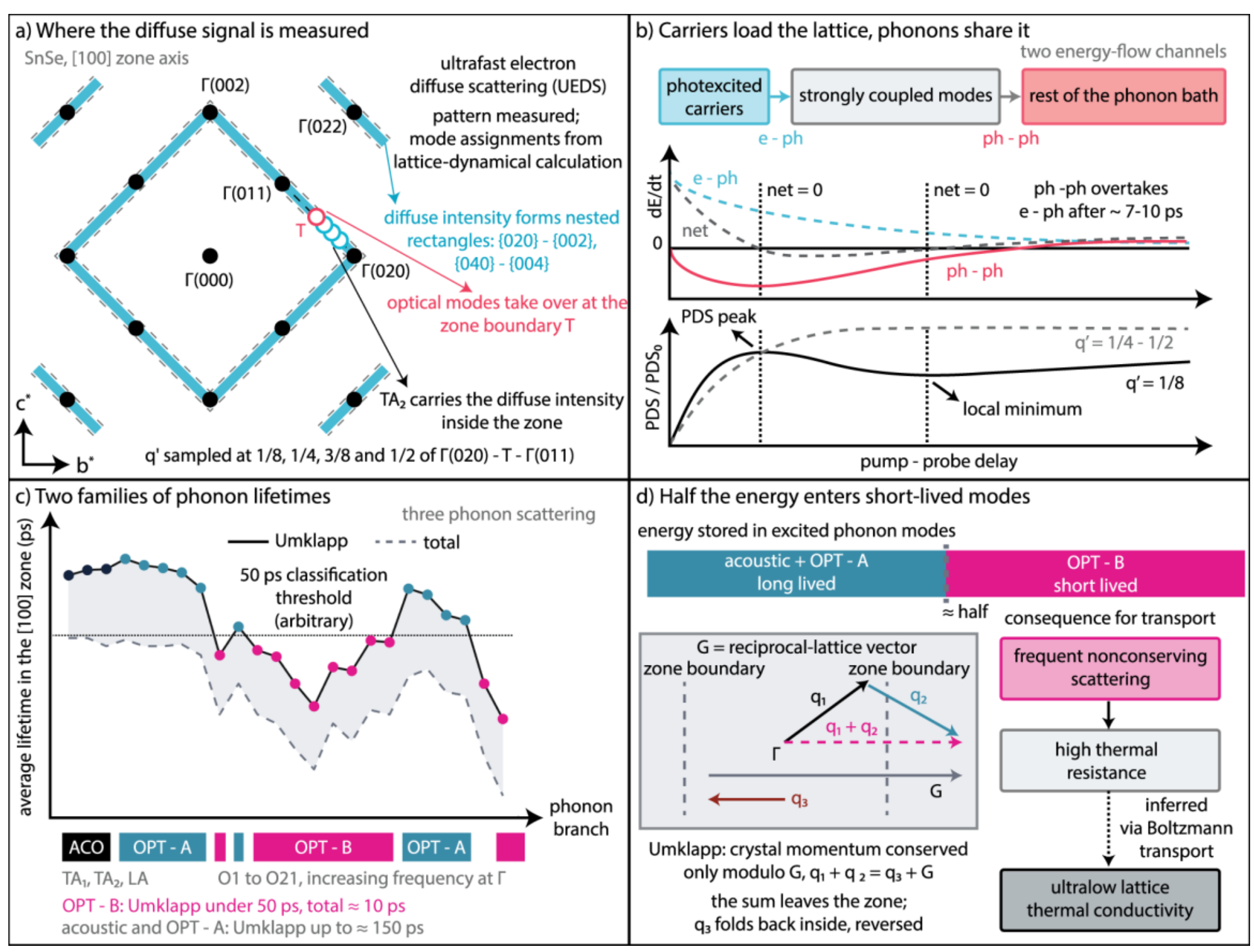

**Figure 5. From nonequilibrium phonons to thermal transport.** (a) Schematic of the ultrafast electron diffuse-scattering signal from SnSe. Diffuse intensity (light blue) forms nested rectangles linking {020} to {002} and {040} to {004}. The transverse acoustic branch $TA_2$ carries most of that intensity; optical modes increasingly dominate toward the boundary point T. Four momentum positions (q') at 1/8, 1/4, 3/8 and 1/2 along Γ(020) to T to Γ(011) sample the crossover. (b) Schematic of the energy cascade pathway following photoexcitation showing the energy-flow channels and relationship between electron-phonon (blue) and phonon-phonon (red) scattering. (c) Schematic of the average phonon lifetimes in the [100] zone: the OPT-B branches, whose total lifetimes are of order 10 ps, and the acoustic branches together with OPT-A, whose Umklapp lifetimes reach roughly 150 ps. (d) Schematic of phonon-branch evolution and relative energy storage. The UEM data revealed that, because Umklapp scattering conserves crystal momentum and only modulates a reciprocal-lattice vector, energy is repeatedly redirected. This gives high

thermal resistance and, through Boltzmann transport, ultralow lattice thermal conductivity. Mode assignments, lifetimes, and the energy partition are deduced from lattice-dynamical calculations and nonequilibrium kinetic theory. Figure content informed by (42).

Phonon excitation and propagation are critical drivers of thermal energy transport, particularly in semiconducting materials (128). UEM imaging uniquely resolves the ultrafast spatial redistribution of coherent strain and phonon-mode amplitude (17), though these observables do not independently provide a calibrated local energy density or heat flux (33). Instead, they reveal how nanoscale features dictate localized energy flow. In $MoS_2$, for example, interlayer strain dynamics precede in-plane strain-wave excitation by 2.4 ps, while step edges as small as a few molecular layers induce observable local phonon dephasing and proximity-dependent softening (99, 129). These in-plane coherent strain waves emerge directly from morphological features, propagate at sonic velocities, and subsequently evolve into broader incoherent structural motion and nanomechanical oscillations – a trajectory consistent with phonon scattering and mode conversion (17).

More broadly, morphology and interfaces actively steer the nucleation, mode selection, propagation, interference, and damping of coherent lattice dynamics. Acoustic wavefronts nucleate at step edges and local strain fields in Ge and $WSe_2$ (16), while terraces or few-layer steps generate spatially distinct oscillation frequencies in $TaS_2$ and $MoS_2$ (98, 124). Further, UEM is highly sensitive to local elastic and bonding properties via changes in phonon velocity dispersion; in GaAs, increasing crystallinity through annealing accelerates the asymptotic coherent-phonon velocity from 4.76 to 5.13 $nm \cdot ps^{-1}$ (88). However, these localized mechanical observations must not be directly equated with bulk thermal conductivity.

Extending beyond coherent phonons, complementary UEM modalities evaluate structural boundaries and charge transport (84, 92, 130–132). For instance, ultrafast CBED and HOLZ-line measurements extract local displacement gradients, strain components, and frequency-wavenumber dispersions in nanostructured semiconducting materials (84, 130, 131). Beyond lattice responses, UEM uniquely captures localized charge and field dynamics: CDEM reconstructs femtosecond photo-Dember charge distributions, whereas electron-optical imaging maps light-induced surface charging and electrostatic potentials in nanomaterials (92, 104, 132). It should be noted that extracting quantitative thermal parameters from these observables requires experimental controls; specimen tilt, translation, dynamical scattering, spatial averaging, nonthermal distributions, long-term drift, irreversible changes, and repetition-rate-dependent heat accumulation should all be accounted for (32–35, 98, 125).

### 4.3. Toward *Operando*, Mode-Resolved Maps of Energy Flow

A compelling frontier for UEM is linking carrier generation and localization to lattice-mode conversion and eventual energy extraction or dissipation within a single, calibrated experiment (16, 41, 82, 109, 133). While individual components of this pathway have been resolved – including exciton-polaron distortions, nonequilibrium phonon thermalization, and defect-launched strain fields – the complete causal sequence remains elusive. To date, no imaging-capable UEM experiment has simultaneously mapped the entire loop of carrier generation, interface localization, lattice-mode conversion, and energy dissipation under authentic device-operating conditions (41, 109, 133, 134).

Correlating real-space imaging, 4D precession electron diffraction, and diffuse scattering would enable UEM to test whether carrier trapping at specific boundaries alters coherent-mode nucleation, propagation, or damping (16, 31, 82, 98, 109, 111, 134). Furthermore, integrating ultrafast EELS, imaging, and near-field measurements could jointly constrain carrier localization, dielectric responses, and branch-selective lattice coupling – provided these observables are spatially registered and independently calibrated (10, 29, 31, 111). Such multimodal integration can successfully distinguish trapping-related electronic contrast from genuine local heating, but only if supported by independent temperature observables, robust controls, and forward modeling.

Currently, this effort is constrained by nonunique SUEM contrast, limited ultrafast EELS energy resolution, spatial delocalization, and the lack of a calibrated modal energy budget (15, 29, 117). While momentum-resolved EELS studies offer a mechanistic precedent, they do not yet provide spatially co-registered real-space imaging (105). To fully transition UEM into a quantitative, statistical platform for mapping interfacial energy flow, these integrated measurements must be implemented under device-relevant boundary conditions, utilizing techniques such as megahertz (MHz) cycling paired with nanosecond thermal recovery (29, 34, 35). Ultimately, sustained, quantitative success in this operational regime will demand strict cycle-to-cycle reproducibility, quantified uncertainty estimates, independent thermometry, dose controls, and comprehensive recovery tests (34, 35, 41).

## 5. Quantum and Correlated Materials: Hidden Order and Competing Degrees of Freedom

While energy materials depend heavily on interfacial transport and thermalization, quantum and correlated materials present a fundamentally different challenge: they host competing electronic, lattice, magnetic, and structural orders that produce transient states that are inaccessible at equilibrium. Because ultrafast excitation couples to these diverse degrees of freedom in distinct ways, it generates metastable states whose true microscopic character must be deciphered by combining complementary contrast mechanisms with structural modeling (10, 12, 41, 48, 54, 135, 136). Furthermore, because these responses are typically spatially heterogeneous, ensemble-averaged measurements often obscure whether an order parameter melts uniformly or through localized domain nucleation, growth, and pinning (12, 13). Dedicated electron-diffraction studies suggest that certain structural order parameters contain distinct excitonic and Peierls contributions; however, isolating theses specific components within UEM generally remains model-dependent (10, 137, 138). Nevertheless, UEM uniquely resolves this nanoscale heterogeneity. In selected platforms, it successfully correlates structural, diffraction, electronic, and field-sensitive proxies directly with local defects and interfaces (key experimental insights are summarized in Table 2) (10, 12, 41, 96, 124, 139).

Table 2. Representative UEM Mechanistic Insights in Quantum and Correlated Materials

| **Material System** | **Excitation/Contrast Mode** | **Key Mechanistic Discovery (Nonequilibrium Pathway)** | **References** |
|---|---|---|---|
| $VO_2$ | PINEM; Energy-filtered UEM; *Operando* UEM | Dielectric metallization occurs in 150 fs before structural lattice response; strained nanoparticles exhibit delayed (0.8 to 1.4 ps) structural induction via strain-associated low-symmetry transient structural phase; electrically driven structural phase-fronts propagate at 4.54 $nm \cdot ns^{-1}$ | (10, 41, 48, 54, 86, 140) |
| 1*T*-$TaS_2$ | Dark-field UEM; 4D UEM; Nanobeam diffraction; *Operando* UEM | Smooth ensemble-averaged CDW decay is driven by binary domain nucleation, 2 ps boundary sharpening, and heterogeneous lateral growth; linear defects launch coherent phonons whose timing and orientation are correlated with subsequent CDW-domain evolution; transition passes through a transient hexatic state | (12, 13, 96, 124, 138, 141) |
| 1*T*-$TaSe_2$ | Precision-controlled UEM | Sub-100-fs CDW structure factor suppression coupled with the activation of multiple-order coherent shear phonons (3.1 THz and 6.2 THz) drives a short-lived metastable stacking reconfiguration | (142, 143) |

| VTe$_2$ | Real-space imaging & 4D-PED | Photoinduced CDW dissolution is accompanied by unusual acoustic-wave generation; the role of shear strain as the triggering mechanism is supported by modeling but not uniquely isolated | (31, 144) |
|---|---|---|---|
| WTe$_2$, MoTe$_2$ | Ultrafast electron diffraction & UEM imaging | Local strain and stacking order redirect structural phase pathways and modify coherent shear or breathing modes; implications for topological phases are model-supported rather than directly measured | (145–148) |
| $\alpha$-RuCl$_3$ | Scanning UEM (SUEM) | Spatially resolved secondary-electron contrast reveals an unusually long-lived hot-carrier transport regime; a pseudospin-dependent cooling bottleneck and photoinduced insulator-to-metal transition are proposed interpretations | (149) |
| FePS$_3$ | UEM nanoimaging | Temperature-dependent changes in coherent shear-mode populations are consistent with magnetic control of phonon scattering; the magnetic mechanism is inferred rather than directly imaged | (150) |

### 5.1. $VO_2$: Real-Space Pathways Through a Correlated Phase Transition

Vanadium dioxide ($VO_2$) serves as an ideal model system for UEM because its insulator-to-metal transition couples electronic, structural, strain, and nanoscale domain responses that can be systematically probed through complementary contrast mechanisms (10, 41, 48, 54, 86). In photoexcited ensembles of $VO_2$ nanoparticles, 4D-UEM and conical-scanning dark-field imaging have revealed particle-dependent diffraction and recovery responses governed by variations in size, orientation, and local strain, rather than a single, uniform ensemble trajectory (140). This observed nonequilibrium contrast is consistent with carrier-induced changes to the interatomic potential and weakened V-V bonding, although the precise transient structure and bond-level mechanisms were not uniquely measured (54, 140). Furthermore, PINEM detected an approximately 155-fs shift in the optical-frequency dielectric response of a single $VO_2$ nanowire

– a change that clearly precedes the structural timescales inferred from prior electron and X-ray measurements (10). Note that this signal must be treated as an electronic-sensitive dielectric proxy rather than a direct measurement of conductivity or simultaneous rutile formation; while the electronic response precedes structural rearrangement, the PINEM measurement itself does not determine the instantaneous crystallographic phase (10).

Energy-filtered UEM has resolved particle-specific structural onset times, revealing that while some particles respond within the ~300-fs experimental resolution, others exhibit delayed responses at 0.8 ± 0.4 ps and 1.4 ± 0.7 ps. Although these delays are consistent with strain-dependent M1-to-M2-to-rutile kinetics, the specific intermediate phases and the complete sequence have not yet been directly resolved (86). Extending these insights into the *operando* regime, UEM mapped a structural phase front proagating at 4.54 $nm \cdot ns^{-1}$ in electrically-driven $VO_2$ devices, a velocity reflecting the structural component of the transition (41). However, when driven at 4 GHz, the device remained predominantly in the rutile state. This incomplete relaxation highlights a device-specific recovery limit governed by heat and phonon-mediated energy dissipation, and therefore should not be interpreted as establishing a universal, intrinsic phonon-limited switching speed for $VO_2$. Figure 6 summarizes key findings of UEM studies of simple optically- and electrically-driven $VO_2$-based devices.

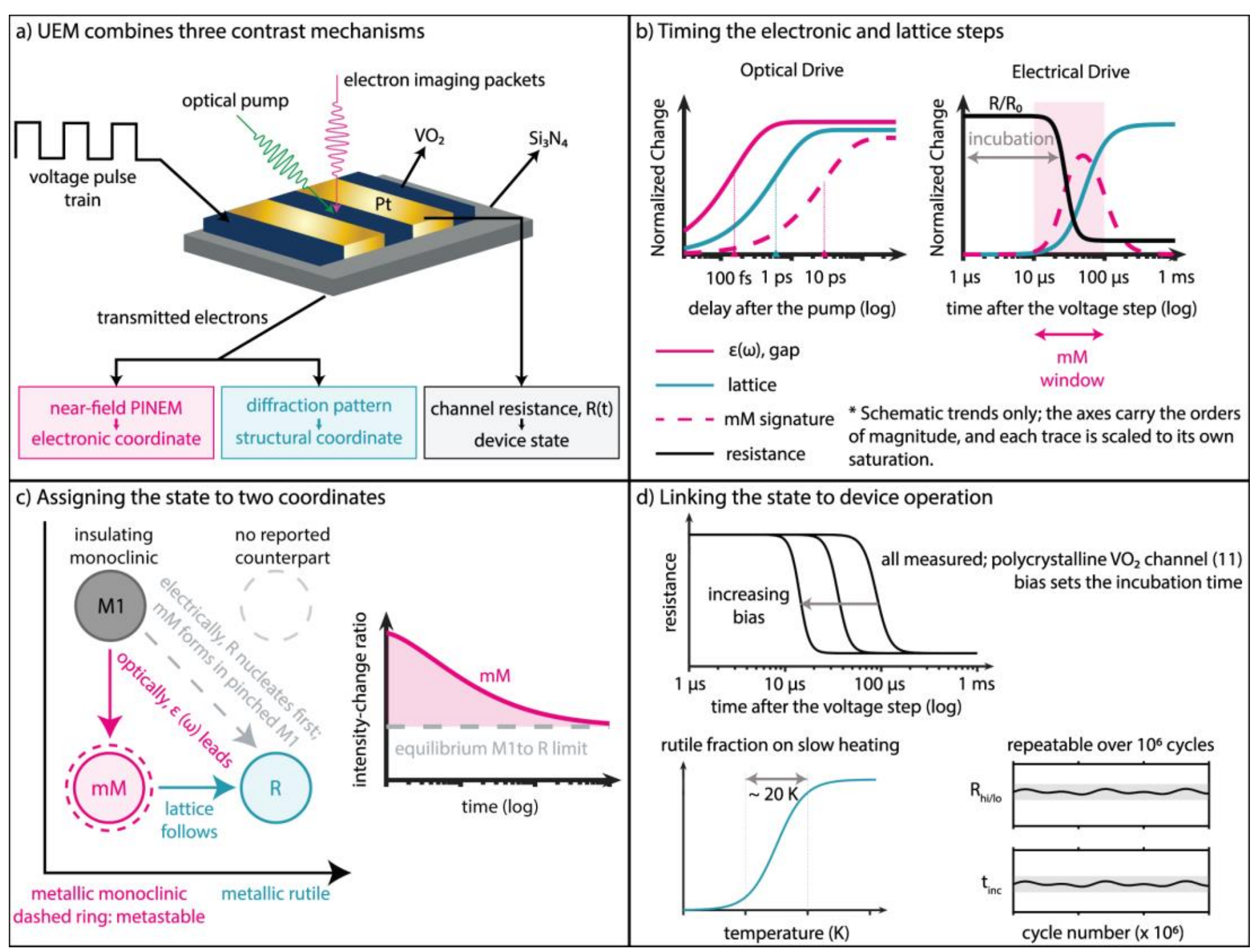

**Figure 6. Electronic and structural pathways in $VO_2$ switching.** (a) Simple schematic of an optically- and electrically-driven $VO_2$ device. By combining UEM modalities, information on electronic and structural coordinates can be gathered. (b) Typical responses of $VO_2$ to the different perturbation types based on PINEM and ultrafast diffraction responses in UEM. (c) Schematic of the model describing the responses, comprised mainly of assignment of different pathways across the structural phase space. (d) Device operation linked to state trajectories associated with the observed incubation time following the applied voltage jump to the simple device shown in (a). Figure content informed by (10, 11, 14).

## 5.2. Phonons as Reporters and Drivers of Hidden Order

In quantum materials, phonons both participate in and provide sensitive structural signatures of competing-order dynamics. UEM directly establishes their generation, propagation, and scattering under nanoscale boundary conditions, linking coherent acoustic dynamics to specific morphological features such as steps, terraces, interfaces, and defects (12, 16, 98, 124, 138). Importantly, UEM can correlate these localized phonon dynamics with the evolution of structural order parameters. In layered metal dichalcogenides, for instance, partially coherent acoustic wave trains gain coherence on a timescale that mirrors incommensurate-domain growth and stabilization (138). Furthermore, UEM has resolved coherent shear dynamics accopanying short-lived stacking reconfigurations (142, 143), and ultrafast nanobeam diffraction has even identified a transient hexatic-like structural state characterized by suppressed translational correlations but retained sixfold orientational order (141). While precision-controlled UEM captures sub-100-fs suppression of the charge-density-wave (CDW) structure factor alongside these multiple-order phonons, it is critical to note that these observations establish coupled structural dynamics rather than strictly phonon-driven switching; whether the phonons actively promote domain evolution or hexatic state formation remains unresolved. Similarly, in $VTe_2$, CDW dissolution coincides with shear-associated lattice deformation and the launch of unusual transverse acoustic waves. Although threshold behaviors suggest a causal link from CDW instability to wave generation, this relationship remains inferential (144).

Coherent phonons and strain fields can also accompany or modulate structural transitions in materials whose stacking configurations are intrinsically linked to topologically relevant electronic states (145–148, 151). In $T_d$-$WTe_2$, for example, wrinkle defects actively redirect photoinduced pathways from a metastable 1T* structure toward a monoclinic 1T' state (145). UEM imaging further resolves standing acoustic waves localized at line defects in $WTe_2$, alongside stacking-dependent breathing and shear modes in octahedral $MoTe_2$ (146, 147). By capturing these dynamics, UEM establishes coherent phonons as highly sensitive probes of local strain and stacking configurations. However, it should be noted that, while these observations provide critical structural input for evaluating topological states, they do not constitute direct measurements of the electronic band structure itself. Finally, UEM extends into magnetic phenomena: in $FePS_3$, for example, the dominant phonon response shifts dramatically across the Néel temperature ($T_N$) – transitioning from interlayer-shear harmonics below $T_N$ to predominantly in-plane propagation above it. This crossover strongly supports a magnetic-order-dependent phonon response, even if the underlying spin-mediated scattering mechanism currently relies on theoretical modeling (150).

5.3. Defect-Pinned Domain Kinetics and Nonthermal Pathways

In quantum materials such as 1*T*-$TaS_2$, domain nucleation, boundary motion, and defect pinning dictate the observed phase trajectories and ultimate device responses (13, 96, 152). Transitioning from qualitative observation to quantitative structural mapping, UEM now directly tracks the evolution of specific order parameters and domain boundaries under optical or electrical excitation (12, 13, 96). By utilizing tailored-aperture ultrafast dark-field imaging to select nearly commensurate CDW satellite reflections, researchers can isolate real-space contrast for the structural order parameter with sub-picosecond temporal and ~5-nm spatial resolution (Figure 7) (12). This capability resolves a critical mechanistic question: it demonstrates that the prolonged ensemble response stems from the heterogeneous lateral growth and recovery of CDW domains – often pinned by local defects – rather than a uniform weakening of the order parameter (13, 152). Indeed, by moving beyond static observations, UEM reveals that linear defects launch coherent acoustic wave trains roughly 100 ps after excitation. While this observation highlights a potential pathway for defect-engineered coupling to CDW dynamics, establishing definitive causal phase stabilization remains an ongoing quantitative measurement objective (138). Finally, UEM *operando* measurements now directly correlate macroscopic resistance features with nanoscale CDW-domain nucleation, capturing abrupt boundary motion near strained regions and dislocations (13, 96). By resolving heterogeneous strain and MHz acoustic responses even during predominantly thermal electrical melting, these studies prove that UEM can now actively parameterize the microstructural realities governing device operation.

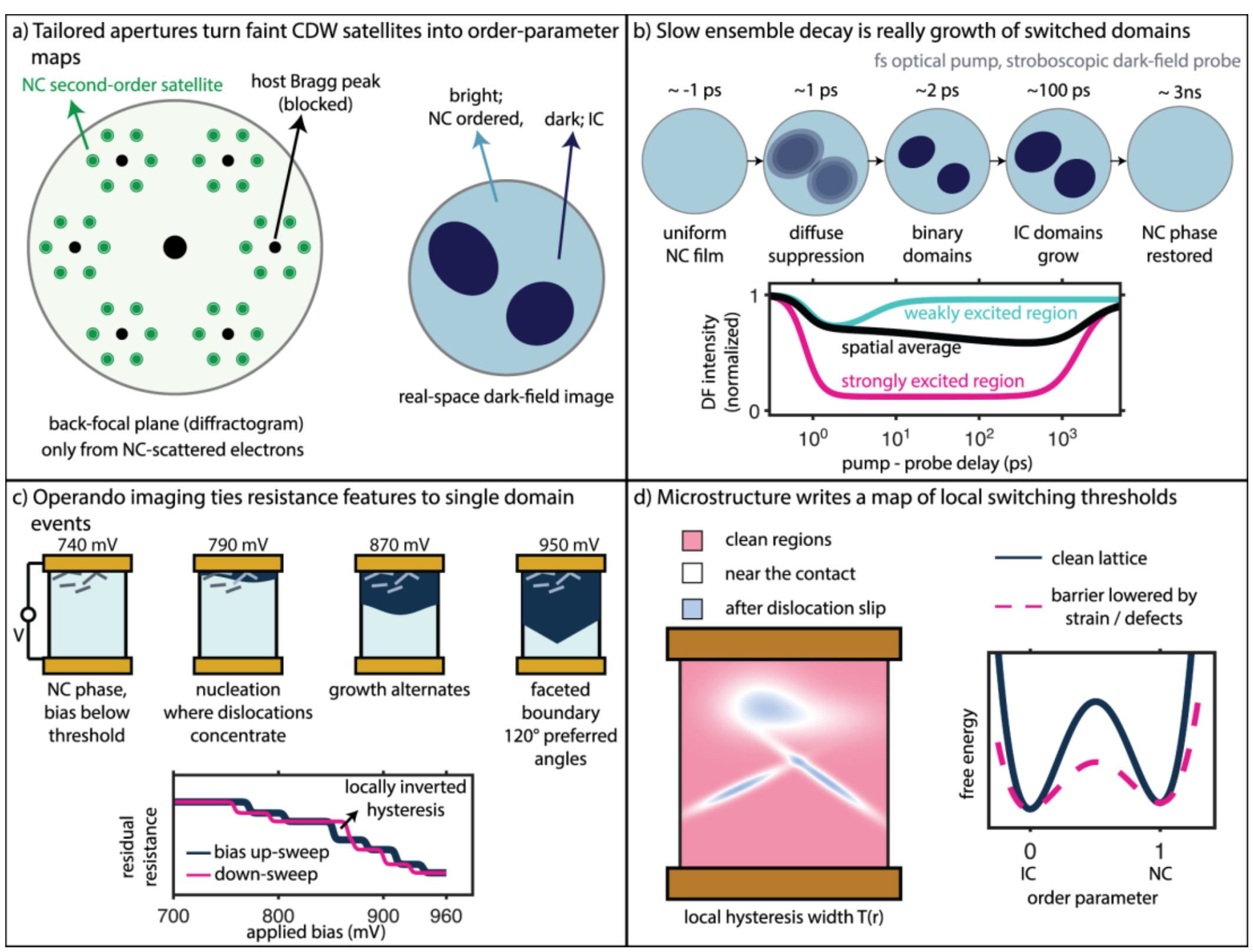

**Figure 7. Spatially heterogeneous phase-transition kinetics.** (a) A tailored aperture array passes only the nearly commensurate (NC) CDW satellites of 1*T*-$TaS_2$ so that dark-field images map the structural order parameter with nanometer-scale and sub-picosecond resolution. Bright regions are NC ordered, while dark regions are incommensurate (IC). (b) After femtosecond optical excitation, a diffuse suppression separates into binary domains, and IC regions grow while domain walls sharpen. Strongly and weakly excited regions follow distinct trajectories, so the slow decay of the spatially averaged signal reflects domain conversion rather than a uniform loss of order-parameter amplitude. (c) Under DC bias, *operando* imaging ties each feature of the residual resistance, including a locally inverted hysteresis, to individual nucleation, pinning, and growth events governed by dislocations, strain, and contacts. (d) Microstructure lowers the local transformation barrier and produces a spatial map of switching thresholds. The macroscopic response is therefore a spatial average over distinct local pathways. Figure content informed by (12, 13).

Optically driven transitions often exhibit electronic-sensitive responses that precede detectable lattice rearrangements, requiring advanced characterization to disentangle nonthermal electronic effects from heating, strain, and phonon dynamics (10, 96, 152, 153). Because carrier redistribution can alter orbital occupations or electronic order before the lattice formally responds, UEM can be used to determine whether subsequent coherent phonons carry the dominant switching energy or merely modulate structural and electronic observables (33, 138, 153). Pushing these capabilities into the spatial domain, SUEM has mapped a long-lived, spatially transported photoexcitation in $\alpha$-$RuCl_3$ over an approximately nanosecond timescale (149). Although assigning this transport to hot carriers and a pseudospin-related cooling bottleneck remains reliant on forward modeling, the measurement underscores UEM's growing utility for parameterizing complex electronic phase spaces. Further exemplifying the shift toward comprehensive multimodal platforms, integrated ultrafast diffraction and EELS in a layered manganite successfully correlated pressure-wave-induced lattice oscillations with simultaneous orbital and electronic spectral changes, identifying rapid heating as the dominant initial driver (153). Indeed, complementary UEM modalities can individually correlate carrier transport, coherent phonons, and structural evolution; integrating them to uniquely determine a complete causal chain stands as a compelling quantitative challenge for the next phase in UEM's growth (10, 138, 149, 153).

## 6. Nanomechanics and Optoelectronics: Local Function, Fields, and Phonons

Just as defects and structural boundaries dictate phase transitions in correlated systems, localized carriers, near fields, and elastic waves govern the nanoscale pathways fundamental to optoelectronic and nanomechanical device function (16, 17, 109, 154–158). Rather than evolving uniformly, photoexcited carriers – alongside directly imaged polariton wave packets and coherent phonons – navigate complex operational landscapes actively shaped by surface termination, band offsets, confinement, structural steps, and strain (16, 18, 109, 158, 159). Because these dynamic processes are inherently morphology-dependent, ensemble-averaged measurements inevitably obscure critical, spatially heterogeneous pathways. Moving beyond qualitative observation, UEM now serves as a predictive characterization platform capable of isolating these specific responses directly at localized interfaces, edges, and individual defects, provided that contrast mechanisms and projection effects are calibrated and controlled (16, 17, 34, 160).

When experimental contrast, geometry, and stroboscopic reproducibility are controlled, UEM transcends qualitative imaging to directly resolve three distinct classes of localized dynamics: anisotropic surface-carrier transport, confined polariton or near-field wave packets, and defect-launched coherent elastic waves. By mapping surface-sensitive photocarrier contrast, SUEM – when coupled with transport modeling – successfully quantifies strong transport anisotropy and interface-mediated variations in effective diffusion (109, 158, 160). Furthermore, PINEM and free-electron interferometry directly visualize confined near fields; here, the analytical rigor is strongest when polariton assignments are directly corroborated by dispersion mapping, phase evolution, or energy-resolved signatures (18, 94, 161, 162). Notably, while UEM directly images coherent elastic wavefronts nucleating or reshaping at localized steps, terraces, interfaces, and defects, translating these observations into quantitative mode assignments requires correlating the data with projection geometries and continuum or atomistic modeling (16, 17, 34, 82, 99).

### 6.1. Spatially Anisotropic Carrier Transport and Surface Potentials

Scanning UEM (SUEM) yields surface-sensitive, real-space spatiotemporal maps of photoexcitation-induced secondary-electron contrast. When integrated with electrostatic and transport models, these measurements transition from qualitative observations to quantitative reporters of hot-carrier motion, localized trapping, charge separation, and surface-potential relaxation (15, 117, 163, 164). More specifically, SUEM resolves contrast signatures indicative of field-driven or Coulomb-mediated charge separation, provided that surface photovoltage, trapping states, and detector responses are systematically calibrated (109, 134, 165, 166). Moving beyond the extraction of effective carrier lifetimes, this approach successfully isolates spatially anisotropic diffusion and interface-directed transport (15, 109, 134, 158). In black phosphorus, for instance, SUEM directly visualized anisotropic surface-contrast expansion, allowing researchers to extract effective hot-hole diffusivities that differed by approximately 15-fold between the armchair and zigzag directions (158). While the resulting diffusion lengths (9.7 and 0.6 μm) remain sensitive to the chosen transport models and temporal windows, such extractions underscore SUEM's capability to derive directional, device-relevant transport metrics directly from nanoscale electron contrast.

By integrating surface-potential measurements with electron-trajectory modeling, SUEM can be used to map the localized electrostatic landscapes that govern carrier motion (15, 117, 133, 164). For example, in GaAs, researchers successfully reproduced surface-termination-dependent dipolar contrast by quantitatively bounding trapping-induced surface potentials at approximately 100 to 200 mV using complementary retarding-field measurements (133). Moving to heterostructures, SUEM visualizations of $MoS_2$ supported on Au demonstrate how thickness-dependent vertical band bending intrinsically generates a lateral potential gradient, actively separating carriers without the need for a chemically distinct semiconductor junction (166). Similarly, at Si/Ge and Schottky interfaces, UEM was used to show that built-in potentials, band offsets, and interface traps decisively redirect hot-carrier motion, often resulting in quasi-two-dimensional interfacial transport. Finally, persistent surface-carrier dynamics in BAs suggest a cooling bottleneck associated with the acoustic-optical phonon frequency gap. Importantly, translating such surface-sensitive observations into definitive models for bulk optoelectronic transport is another compelling opportunity for future advancement of the UEM platform generally (109, 117, 134, 160, 165).

### 6.2. Confined Polariton Wave Packets

Free-electron methods provide a combination of nanometer-scale spatial resolution, broadband momentum coupling, and high sensitivity to confined optical fields (167–169). Advancing beyond the measurement of steady-state dispersion relations, PINEM now captures quantitative spatiotemporal maps of phonon-polaritons in media such as hexagonal boron nitride (h-BN) (18). In optimized pump-probe UEM geometries, energy-filtered PINEM successfully maps the interaction-weighted evolution of polariton wave packets, provided that critical experimental parameters – such as electron-pulse duration, temporal matching, and wave-packet group velocity – are controlled (170). These advanced measurements reveal how group velocities evolve during propagation, capturing the wave packet splitting into spatially separated components. However, physical interpretation requires acknowledging that this spatial splitting does not inherently indicate true mode conversion or the spontaneous creation of independent polariton eigenstates. Nevertheless, because electron probes in selected transmission geometries are sensitive to buried or interface-bound modes (167, 170), this capability elevates UEM into a metrology platform for engineering pulse shaping, routing, and interference within confined two-dimensional heterostructures and next-generation on-chip photonic systems (18, 94, 170).

### 6.3. Defect-Launched Elastic Waves

Transitioning into the nanomechanical regime, the time-varying, morphology-dependent contrast in UEM images can be leveraged to go beyond qualitative observation, actively mapping the defect-associated nucleation and modulation of coherent elastic waves (16, 17, 33, 34). By integrating tracked spatiotemporal displacements with mechanical models, the platform now routinely extracts effective elastic parameters, resonant frequencies, damping times, and localized force-related quantities directly from individual nanomechanical architectures (59, 171). This quantitative capability represents a significant maturation from the field's foundational proof-of-concept demonstrations – such as early observations in suspended graphite films that captured the macroscopic transition from complex mechanical modes into a dominant 1.08-MHz global drum resonance – by shifting the focus toward predictive extraction of mechanical metrics in complex, localized geometries (172).

Advancing well beyond the observation of global resonances, multimodal UEM now quantitatively maps how localized structural discontinuities launch and modulate the direction, mode content, and attenuation of propagating coherent elastic waves (16, 17, 173). Through real-space imaging, UEM can be used to capture acoustic wavefronts nucleating directly at step edges, bend contours, and pre-existing strain fields in materials such as Ge, $WSe_2$, and $MoS_2$. Rather than simply observing these waves propagate, their subsequent mode-selective damping, dephasing, interference, and coupling can be tracked, though invoking specific phonon-phonon scattering mechanisms typically requires corroboration by kinetic data and forward modeling (17, 33, 98). Pushing these analytical capabilities further, ultrafast CBED measurements in graphite have exposed the limits of classical elasticity; by successfully extracting time-dependent expansion and shear-rotation components, UEM has been used to provide localized empirical data required to accurately model ballistic energy transport, nonequilibrium phonon populations, and microscopic dissipation (82). Ultimately, these measurements transition nanomechanics from a macroscopic continuum assumption into a predictable, localized structure-dynamics problem. By directly correlating morphological features – such as terraces, steps, and polymer chain orientations – with quantitative changes in the phase velocity, resonance, and relaxation of coherent acoustic phonons, UEM can now be used to establish a metrological foundation for engineering active mechanical nanostructures (84, 88, 174–176).

Presently, extracting quantitative metrics from bright-field UEM contrast requires forward modeling. Researchers must systematically distinguish true structural displacement from a host of confounding signals, including excitation error, dynamical scattering artifacts, variations in specimen orientation and local thickness, beam drift, and background evolution (31, 33, 34). By successfully isolating the precise sources and trajectories of coherent modes alongside diffraction-based phonon population measurements, UEM now provides tools for actively characterizing and tailoring coherent acoustic sources, cavities, and mode distributions within selected geometries (150, 173, 177). While recent studies of patterned bilayers and nanomechanical cavities have successfully demonstrated this controlled excitation and precise mode shaping, the next major frontier for the field lies in expanding these targeted capabilities into a generalized metrology for engineering active, programmable mechanical waveguides. The three examples given in this section are summarized in Figure 8.

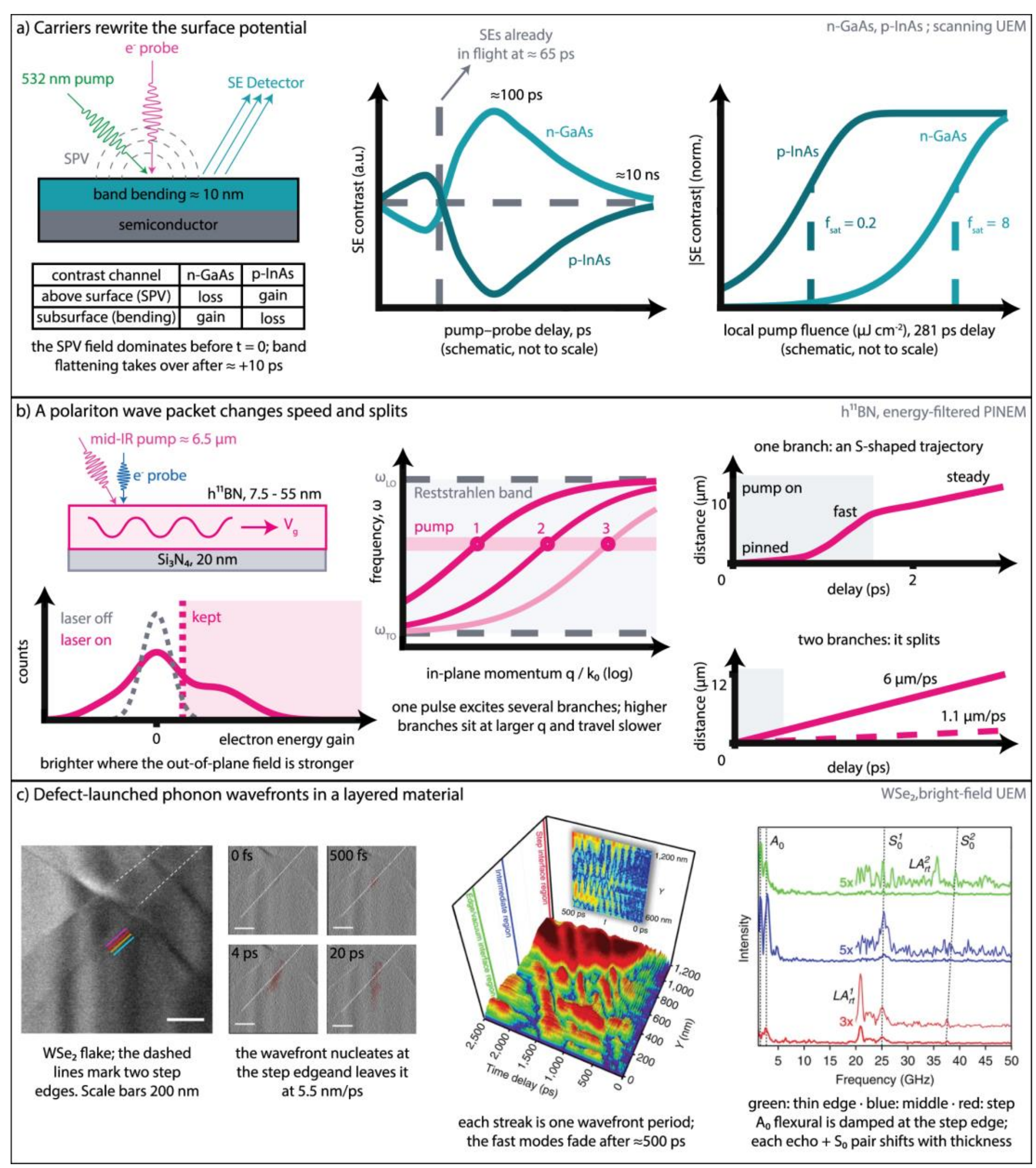

**Figure 8. UEM visualizes propagating excitations at device-relevant length scales.** (a) SUEM maps the photoinduced surface potential, not the carrier density. An above-surface photovoltage channel acts before $t$ = 0 and a subsurface band flattening channel after it, and because the two carry opposite signs the contrast inverts across $t$ = 0. The implied surface state densities, $5 \times 10^{11}$ and $2 \times 10^{13}$ cm$^{-2}$, differ by a factor of 40, yet both materials peak near 100 ps. (b) Energy filtered PINEM follows a phonon polariton packet in h$^{11}$BN. Hyperbolic dispersion places several branches within one pump bandwidth; higher branches sit at larger momentum and travel more slowly, from 0.35 to 6.7 $\mu m \cdot ps^{-1}$. A packet is pinned at the excited edge, accelerates, then settles, and a multibranch packet separates at 6 and 1.1 $\mu m \cdot ps^{-1}$, indicating two branches excited together rather than conversion of a single mode. (c) A wavefront nucleates at a $WSe_2$ step edge and launches at 5.5 $nm \cdot ps^{-1}$, with each streak marking one wavefront. In each subregion the dominant pair is the *c*-axis acoustic echo and a dilatational Lamb mode, with frequencies depending upon local flake thickness. The flexural $A_0$ branch is damped at the step. Geometry, defects, and interfaces set where energy starts and how it travels, but carriers spread diffusively where polaritons and phonons propagate as waves. Content in (a) and (b) informed by (15, 18); panel (c) adapted from (16).

### 6.4. Brief Perspective on UEM in Optoelectronics and Nanomechanics

UEM quantitatively demonstrates how localized carrier, polariton, strain, and domain trajectories inherently constrain macroscopic functional behavior through interfaces, defects, and thermal pathways (13, 41, 109). As the platform matures, an ongoing challenge lies in correlating complementary mechanical, electronic, and chemical observables across unified UEM workflows – ideally within the exact same operating device (28, 29, 48, 95). Achieving this level of multimodal integration is likely to uncover new physical insights, such as directly mapping how transient acoustic deformations uniquely correlate with, or actively drive, localized changes in electronic structure. While complementary studies across the field currently supply individual links within the broader excitation-transport-strain-dissipation chain, the platform's next evolutionary leap is clear. Continued development is actively driving the transition of UEM from merely observing heterogeneous intermediate states into a metrology platform fully capable of parameterizing mechanistic models of device operation and identifying experimentally testable physical limits (13, 18, 29, 35, 41, 82, 109).

## 7. From Specialized Experiments to Broad-Ranging Materials Platform

The initial phase of UEM successfully synchronized pulsed electron sources with optical or electrical excitation to probe nonequilibrium materials phenomena (28, 50, 52, 53, 80, 178). Building on this foundation, the field is now transitioning from specialized, proof-of-concept demonstrations into a reliable, quantitative materials platform (40). To achieve this metrological standard, focus is shifting toward rectifying key barriers to quantifiability. This includes operating at the extremes of dose limits, systematically minimizing instrument-induced artifacts, and incorporating dynamical scattering – which must be quantitatively bounded under strict specimen-, thickness-, orientation-, and energy-specific conditions (27, 31, 82). Because confounding artifacts stem from myriad sources – such as sample drift, beam fluctuations, tilt changes, morphology evolution, damage, detector saturation, and experimenter bias (34, 37) – a true platform-level measurement should ideally report calibrated uncertainty, dose and detector responses, repeatability across specimens, forward-model validation, and independent artifact controls (27, 34, 37, 179, 180). Furthermore, UEM maturation will continue to expand into advancements in *operando* excitation, executing measurements under functional electrical, thermal, chemical, or mechanical boundary conditions while simultaneously monitoring the relevant macroscopic device response (11, 13, 41). Ultimately, cementing UEM as a definitive quantitative tool will most certainly rely on validated multimodal registration, quantitative contrast models, and strict statistical controls to separate genuine physical dynamics from long-term acquisition artifacts (28, 34, 95).

7.1. *Operando* Excitation and Device Relevance

The most definitive route to establishing true macroscopic device relevance is to interrogate specimens under authentic electrical, thermal, optical, or mechanical boundary conditions (11, 13, 41). While traditional optical excitation successfully accesses related metastable phase pathways, it is fundamentally not meant to impose electrode constraints, localized current-density distributions, contact resistances, Joule-heating profiles, and substrate-mediated strains inherent to actual electrical operation (11, 41, 81). Accordingly, as UEM matures into a predictive engineering tool, *operando* workflows will integrate these realistic boundary conditions to accurately parameterize active device function.

Indeed, *Operando* UEM of electrically driven $TaS_2$ has successfully correlated nanoscale CDW-domain nucleation, growth, defect pinning, and phase-front motion with macroscopic resistance changes, mapping bend-contour evolution and spatially heterogeneous responses in real time (13, 96). Again, moving beyond visualization for visualization's sake, these electrically driven measurements resolve phase-front propagation to identify fundamental repetition-rate limits governed by thermal and phonon-mediated structural relaxation (11, 41). In $VO_2$, for example, UEM has been used to demonstrate that structural recovery and reversible switching are dictated by heat dissipation – revealing device-specific limits dependent on film thickness, substrate architecture, contact geometry, and duty cycle (35, 41, 126). Consequently, to serve as a truly predictive metrology tool, next-generation *operando* UEM platforms will natively combine active electrical biasing with simultaneous transport readouts, localized field or temperature diagnostics, and thermal-management controls.

7.2. Energy-Resolved and Chemically Specific Contrast

While ultrafast diffraction and real-space imaging effectively capture structural dynamics, UEM-based energy-resolved techniques – such as ultrafast EELS, PINEM, and CDEM – quantitatively constrain underlying electronic, vibrational, and chemical mechanisms (29, 38, 46). Translating these spectral signatures into definitive material responses requires instrument calibration coupled with electrodynamic or electronic-structure modeling. Highlighting this quantitative maturation, laser-scanned PINEM has successfully achieved ~20-meV resonance mapping for optically accessible plasmonic modes, managing the fundamental spectral-temporal trade-off dictated by the laser bandwidth (181). Similarly, energy-filtered UEM has been used to map strain-sensitive, particle-dependent phase-transition pathways in $VO_2$ with ~700-fs temporal resolution (86). Furthermore, while CDEM maps electron energy shifts, beam deflections, and THz near fields, researchers now reconstruct precise spatiotemporal maps of carrier density by integrating such empirical observables with hydrodynamic and electrostatic models. Pushing analytical sensitivities even further, electron-interference measurements – including Ramsey, homodyne, and Lorentz-PINEM methods – now successfully retrieve weak near-field amplitudes and phases, thus expanding UEM's capacity for comprehensive electronic characterization (92, 94, 106, 169).

Ultimately, these methods transform phase- and amplitude-sensitive electron signals into quantitative reconstructions of near-field and charge-dynamics observables (18, 92, 94). To extend this predictive power into the chemical domain, inherently weak core-loss cross sections and low-loss spatial delocalization will need to be overcome. Achieving this demands next-generation instrumentation – for example, seamless integration of high-coherence field-emission sources, lossless monochromation, and direct electron detectors (29, 46, 182–184). Furthermore, as with conventional TEM, core-loss edges provide element- and valence-sensitive contrast, and UEM frameworks will need to incorporate spectral deconvolution against established reference spectra. This will further require corroboration by electronic-structure or multiple-scattering calculations to ensure uniquely determined chemical assignments (38, 46, 182).

### 7.3. Statistics, Repetition Rate, and Sample Recovery

High-resolution UEM inherently involves navigating complex, coupled tradeoffs among temporal resolution, electron number per pulse, signal-to-noise ratio, and space-charge broadening (185). For beam-sensitive materials, the integration of direct electron detectors significantly improves dose efficiency. However, to maintain quantitative rigor, their implementation must account for pileup, dynamic range limitations, and cumulative sample damage (37, 179, 186, 187). Mapping reversible dynamics at high repetition rates demands that the specimen reaches a strictly reproducible cycle-to-cycle state – either by fully returning to its true ground state within experimental uncertainty or by entering a characterized periodic steady state that accounts for residual heat, strain, and phase memory (11, 35, 41). Demonstrating this platform-level control, MHz-scale data acquisition is now feasible via thermally engineered specimen supports; achieving 2-MHz cycling in a model CDW geometry, for example, required extracting nanosecond-scale relaxation parameters directly from coupled structural recovery and heat-flow modeling (35). Ultimately, to definitively separate true excitation-driven physics from long-term drift, beam-induced artifacts, or experimenter bias, the field will ideally adopt standardized statistical validation protocols. Moving forward, a robust UEM reproducibility framework should include randomized acquisition ordering, automated drift tracking, objective signal-quality clustering, detector calibration, and strict dose controls, culminating in comprehensive uncertainty propagation and independent replication (27, 34, 37, 179).

7.4. Four- and Five-Dimensional Acquisition

Advancing into four- and five-dimensional (4D and 5D, respectively) acquisition fundamentally alleviates persistent ambiguities related to excitation error, orientation, projection, and mode overlap, provided these additional dimensions are calibrated and modeled (27, 31, 188, 189). For instance, ultrafast 4D precession electron diffraction (4D-PED) averages over incident angles to suppress dynamical-scattering artifacts, revealing true quantitative refinement of thickness, symmetry, and atomic coordinates under specified assumptions (31). Pushing these boundaries further, ultrafast CBED and subsequent 5D-STEM-CBED implementations now map strain tensors across multiple real-space dimensions with nanoscale and sub-picosecond to picosecond precision (82, 84). By extracting strain components from CBED-line shifts, these techniques enable comprehensive mapping of acoustic wave amplitudes, polarizations, and momentum-frequency dispersions – relying on elastodynamic calculations and Fourier analysis to ensure uniquely determined mode assignments. Complementing diffraction, bright-field UEM isolates localized, low-frequency (*e.g.*, ~30-MHz) components – such as defect-associated vibrations near physical edges – by extracting signals from pixelwise Fourier spectra and corroborating them against global-mode modeling (85). Importantly, scaling into these higher dimensions increases data volume and acquisition time, elevating the metrological stakes; extracting reliable physics from 4D and 5D datasets requires calibration of angular registration, beam position, sample drift, detector response, and cumulative dose (34, 37, 48, 188).

7.5. Validation and Quantitative Inference

The hallmark of UEM's maturation is the recognition that raw experimental contrast does not automatically yield quantitative physical parameters (15, 33, 190). Because observed contrast is intricately linked to complex experimental realities – ranging from dynamical scattering, excitation errors, and local thickness variations in oriented crystals, to surface band bending and trapping states in SUEM, or specific electron-field interactions in PINEM and CDEM – forward modeling is presently a requirement. For example, applying simplified kinematical models to bright-field phonon contrast in strongly scattering crystals can result in substantial underestimations or mode misassignments. Conversely, quantitative UEM becomes more straightforward when extracting localized strain via CBED, provided that the requisite kinematical conditions are strictly calibrated and validated (33, 82, 84, 130). Furthermore, because thickness-dependent diffraction backgrounds, evolving bend contours, and time-dependent local orientations can easily generate the illusion of wave propagation, isolating simple Fourier peaks is not sufficient to establish true phonon dispersion. Instead, genuine structural dispersions are authenticated only by directly correlating diffraction data with elastic simulations and thickness controls, conclusively separating transient materials physics from dynamic background artifacts (16, 33, 82, 84, 130).

Accordingly, achieving true quantitative inference dictates that the relationship between the detector signal and the underlying physical mechanism be modeled, utilizing dynamical-scattering simulations, refined structural coordinates, and modality-appropriate forward models. UEM is indeed rapidly progressing in this direction. For example, while dynamical multislice or 4D-PED refinements are necessary for strongly diffracting crystals (27, 31), geometric CBED models suffice in validated weak-scattering regimes (82). Relatedly, EELS, PINEM, and CDEM require sophisticated spectral or electrodynamic frameworks to properly extract intrinsic materials properties (92, 126, 169).

Indeed, the most authoritative methodologies now triangulate complementary observables – such as correlating structural diffraction with energy-loss spectroscopy – while embedding experimental controls like fluence scaling and dose tracking (86, 96). These multimodal approaches achieve their highest predictive power when data are acquired under identical, co-registered conditions and independently constrain distinct facets of the physical behavior. For instance, while pairing structural diffraction with EELS or optical reflectivity provides valuable insight, agreement does not guarantee a uniquely determined inversion if both analyses rely on shared model assumptions (46, 105, 191). Because fluence scaling and electron-dose tracking alone cannot definitively exclude long-term drift, dynamic tilt, irreversible material degradation, or experimenter bias (34, 84, 126), platform-level validation requires a stricter standard. Thus, to fully complete its transition into a mature characterization tool, UEM workflows should systematically enforce comprehensive controls: orientation and thickness series, baseline static reference states, active drift and tilt monitoring, verified reversibility, randomized acquisition ordering, exhaustive uncertainty propagation, and beam-damage checks.

## 8. Outlook: UEM as a Quantitative Materials Instrument

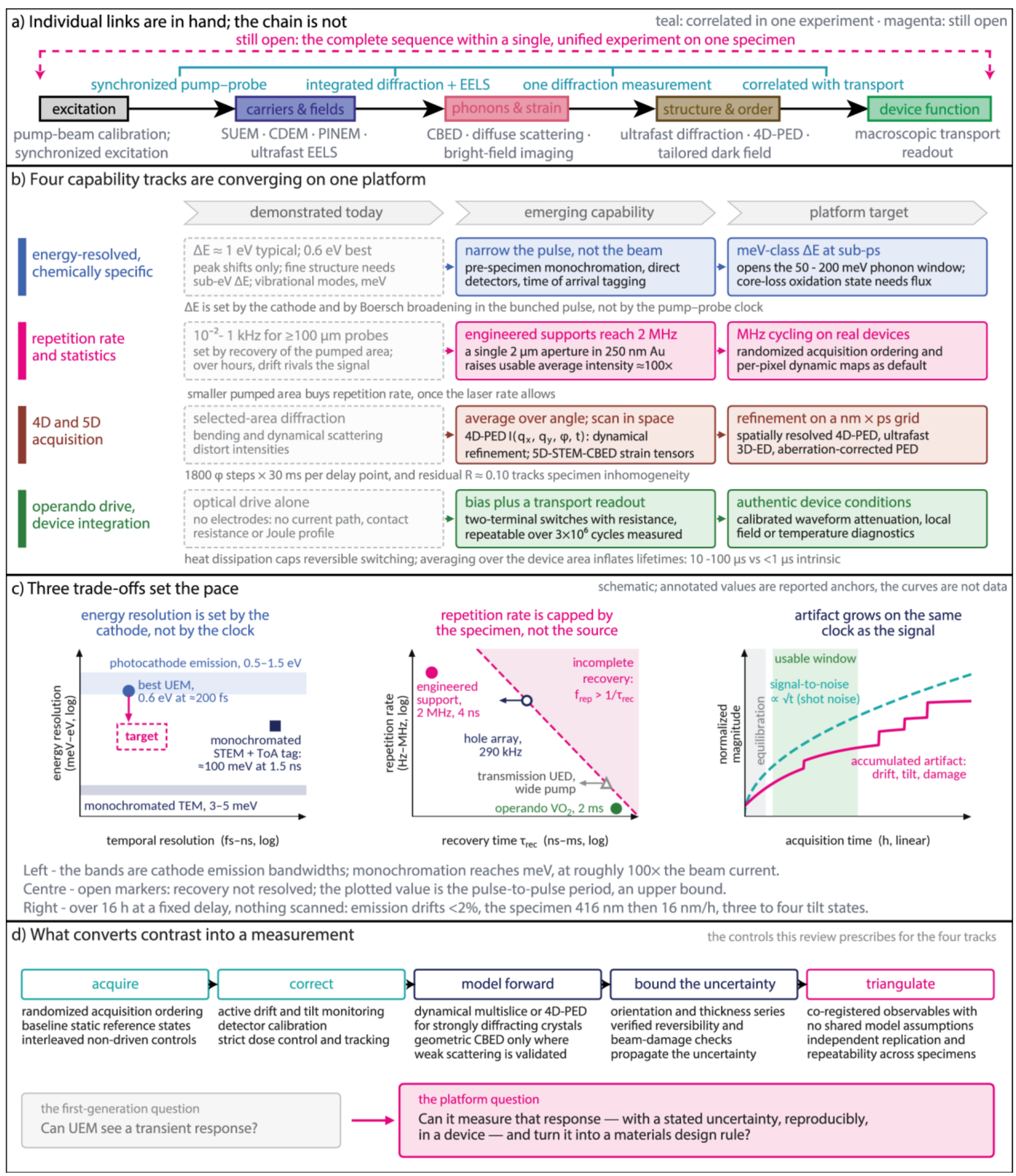

**Figure 9. From proof-of-concept to a quantitative materials platform.** (a) The causal sequence UEM must resolve, with the modality reporting each stage. Every adjacent pair has been correlated in one experiment (teal); the complete sequence within a single, unified experiment on one specimen has not (magenta). (b) Four capability tracks are converging: energy-resolved detection, repetition rate and statistics, four- and five-dimensional acquisition, and operando drive. Each runs from what is routine today, through what is emerging, to what the platform requires, and closes with the limit that track does not by itself remove. (c) Three trade-offs set the pace: energy resolution is set by the photocathode and by Boersch broadening rather than by the pump-probe clock; repetition rate is capped by recovery of the pumped area rather than by the source; and artifacts accumulate over the same integration that improves signal-to-noise. Axes are schematic and the curves are not data; annotated values are reported anchors, and open markers mark unresolved recovery times, for which the plotted value is the pulse-to-pulse period. (d) The controls that convert contrast into a measurement, and the transition the field now faces: from asking whether UEM can see a transient response to asking whether it can measure that response (with a stated uncertainty, reproducibly, in a device) and turn it into a materials design rule. Figure contents informed by (11, 13, 29, 31, 34, 192).

UEM is entering a transformative new phase. While the first phase successfully established that synchronized pulsed electron sources could probe nonequilibrium materials, the next phase will be defined by quantitative, reproducible, and device-relevant metrology (Figure 9). To achieve this, the instrument will evolve into a fully integrated, multimodal platform. While imaging, diffraction, holography, Lorentz microscopy, and spectroscopy have all been demonstrated on high-coherence UEMs, they generally operate as separate or sequential modes (23, 44). Integrating these structural techniques with ultrafast EELS, PINEM, CDEM, and other coherent near-field methods will provide holistic sensitivity to carriers, phonons, plasmons, fields, and bonding – observables that are currently measured piecemeal (29, 92–95). For example, although ultrafast diffraction and EELS have been correlated in carbon nanotubes, the measurements were performed sequentially on different instruments (91); similarly, SUEM and cathodoluminescence probe complementary sample depths but suffer from mismatched temporal resolutions (160). Thus, a defining objective for the next generation is to map a complete causal sequence simultaneously within a single specimen: from initial excitation to carrier or field motion, through subsequent phonon and strain responses, and finally to local structural evolution and macroscopic device function (17, 86, 94, 109, 166). Indeed, closing this gap is a compelling direction, as no current demonstration has yet established the complete excitation → carrier → phonon → structure → device-function sequence within a single, unified UEM experiment (41, 48).

A second defining trajectory for the platform is the incorporation of *operando* environments. While targeted *operando* studies have begun successfully linking electrical stimuli to structural domain evolution and macroscopic transport, for example, the simultaneous measurement of localized fields, carrier populations, transient temperatures, and phonon dynamics remains an outstanding challenge (13, 41, 96). Capturing realistic device operation requires calibration of electrical waveform attenuation, pump-beam parameters, thermal recovery rates, electron dose, and contact geometries (35, 180, 193). Furthermore, pushing to higher repetition rates does not guarantee baseline recovery; as seen in $VO_2$, structural switching becomes strictly recovery-limited at GHz frequencies (41). Consequently, future UEM platforms may be designed around authentic device architectures, actively integrating realistic thermal pathways and electrical contacts to quantitatively interrogate switching barriers, failure nucleation, and carrier extraction under true operating conditions.

Quantitative, multidimensional acquisition represents a third evolution. Methods such as ultrafast CBED, 5D-STEM, and 4D-PED now enable nanoscale metrology in calibrated regimes to extract local strain tensors, acoustic-wave polarizations, structural dispersions, and nonequilibrium atomic coordinates (31, 82, 84, 139). However, converting such rich observables into actionable parameters for atomistic and continuum modeling requires forward modeling, accounting for dynamical-scattering calculations, and propagating uncertainty (33). Underpinning all of this, statistical reliability will ideally become a foundational UEM capability. Whether deploying high-repetition-rate measurements on thermally engineered supports (35) or low-repetition-rate direct electron detection for dose-sensitive conditions (22, 186), experiments should be universally coupled with statistical validation protocols. By making randomized acquisition ordering, automated drift correction, and objective signal clustering the norm, the field can significantly reduce experimenter bias and beam instabilities with the aim of definitively separating weak but genuine physical signals from systemic instrumental artifacts (34).

The ultimate scientific consequence of these integrated capabilities is a profound paradigm shift in nonequilibrium materials science and engineering. Rather than passively observing homogeneous bulk dynamics, researchers are now actively characterizing – and in selected systems, deliberately patterning – defects, interfaces, grain boundaries, and strain gradients to modify nonequilibrium pathways. These morphological features are no longer viewed merely as scattering sites; they are being engineered as active elements to deliberately launch, guide, trap, or dissipate transient energy (16, 84, 99).

While stroboscopic UEM is uniquely suited for mapping reversible transients, exploiting such defects, interfaces, and grain boundaries to modify nonequilibrium pathways must consider possible dissipative effects. Importantly, attentuation or mode conversion, for example, does not by itself prove localized dissipation: elastic redistribution, interference, phonon-phonon scattering, carrier trapping, dephasing, and thermalization must be distinguished through controls and modeling (33, 34). Repeated excitation can remain reversible when thermal and structural recovery are adequate, as demonstrated by MHz cycling with engineered heat dissipation (35). However, insufficient recovery, beam dose, charging, drift, or cumulative heating can produce dissipation that drives irreversible phase evolution or structural degradation. Importantly, to fully capture irreversible processes, stroboscopic UEM can be complemented with single-shot or movie-mode dynamic TEM (DTEM), albeit with trade-offs in spatial resolution, dose, space-charge broadening, frame depth, and statistical repeatability (72, 74, 75, 187, 194). Together, stroboscopic UEM and DTEM can connect repeatable early-time dynamics to selected irreversible transformations, but a holistic causal account requires matched excitation conditions and correlative measurements rather than sole concatentation of the two modalities.

While the pursuit of attosecond and few-femtosecond temporal resolution remains essential for isolating specific optical-cycle, charge-migration, and nonadiabatic field-driven processes, such resolutions are often less relevant for addressing the structural and *operando* questions governing device function (106, 195, 196). Ultimately, temporal resolution must scale appropriately with the targeted materials physics. For the vast majority of functional systems, the greater opportunity lies not in pushing for better temporal resolution, but in combining sufficient sub-picosecond timing with chemical specificity, authentic *operando* environments, and quantitative spatial interpretation.

Ultimately, we argue that the central design rule for the field is clear: UEM provides its highest value when a decisive macroscopic materials property emerges directly from a transient, localized, and highly coupled nanoscale pathway. By quantitatively identifying these pathways, definitively distinguishing between competing physical mechanisms, and revealing precisely which structural or instrumental variables can be engineered, UEM will shed its historical identity as merely a “faster electron microscope”. Instead, it will continue to grow into an indispensable, predictive metrology platform for designing and characterizing advanced materials in functional states that equilibrium thermodynamics simply cannot describe.

Summary Points:

1. A Metrological Paradigm Shift: Ultrafast electron microscopy (UEM) has matured from a qualitative, proof-of-concept imaging tool into a rigorous, predictive metrology platform for nonequilibrium materials engineering.
2. The Complete Causal Sequence: Establishing the full mechanistic chain of materials function – from initial excitation to carrier dynamics, phonon/strain responses, structural evolution, and ultimate device function – requires measuring these observables simultaneously rather than sequentially.
3. Multidimensional Quantification: The integration of 4D and 5D acquisition methods (such as ultrafast CBED, 4D-PED, and 5D-STEM) empowers the extraction of quantitative local strain tensors, acoustic-wave polarizations, and nonequilibrium atomic coordinates.
4. Strict Analytical Standards: Translating transient electron contrast into definitive physical parameters demands robust forward modeling, dynamical-scattering calculations, and objective uncertainty propagation, unequivocally replacing simplified kinematical assumptions.
5. Statistical Reliability as a Foundation: Distinguishing weak, genuine physical signals from systemic instrumental artifacts necessitates universally adopted validation protocols, including randomized acquisition ordering, active drift correction, and rigorous dose controls.

6. Engineering Localized Pathways: UEM uniquely reveals how defects, interfaces, grain boundaries, and strain gradients function not merely as scattering sites, but as active, patternable elements that can deliberately guide, trap, or dissipate transient energy.

Future Issues:

1. Achieving Holistic Multimodal Integration: While imaging, diffraction, holography, and energy-resolved spectroscopies are mature standalone modes, the field must develop unified platforms capable of co-registering these observables simultaneously within a single specimen.
2. Authentic Operando Architectures: Future platforms must be intrinsically designed around realistic device geometries, integrating authentic electrical contacts, localized field diagnostics, and realistic thermal pathways to probe switching barriers under native operating conditions.
3. Balancing Resolution with Relevance: While extreme (attosecond) temporal resolution serves specific field-driven processes, the far broader challenge for functional systems lies in combining sufficient sub-picosecond timing with robust chemical specificity and localized spatial interpretation.
4. Standardized Reproducibility Frameworks: The community must coalesce around standardized, open-source computational frameworks for inverse modeling and statistical clustering to eliminate experimenter bias and ensure cross-laboratory reproducibility of multidimensional datasets.

Disclosure Statement

The authors are not aware of any affiliations, memberships, funding, or financial holdings that might be perceived as affecting the objectivity of this review.

Acknowledgments

This material is based on work supported by the U.S. Department of Energy, Office of Science, Office of Basic Energy Sciences under Award No. DE-SC0023708.